\documentclass[12pt]{article}
\usepackage{tocloft}
\usepackage{CJKutf8}
\usepackage{amsfonts}
\usepackage{amsmath}
\usepackage{amssymb}
\usepackage{array}
\usepackage{bigints}
\usepackage{bm}
\usepackage{booktabs}
\usepackage[nosort]{cite}
\usepackage{color}
\usepackage{dsfont}
\usepackage{float}
\usepackage{framed}
\usepackage{graphicx}
\usepackage{indentfirst}
\usepackage{mathrsfs}
\usepackage{multirow}
\usepackage{pdflscape}
\usepackage{setspace}
\usepackage{subdepth}
\usepackage{subfig}
\usepackage{titlesec}
\usepackage{wrapfig}
\usepackage[all]{xy}
\usepackage{young}
\usepackage[vcentermath]{youngtab}
\usepackage{relsize}
\usepackage{stackengine}
\usepackage{verbatim}
\usepackage{slashed}
\usepackage[utf8]{inputenc}
\usepackage{enumitem}

\usepackage{hyperref}
\hypersetup{colorlinks=true}
\hypersetup{linkcolor=black}
\hypersetup{citecolor=black}
\hypersetup{urlcolor=black}

\numberwithin{equation}{section}

\usepackage[left=2.5cm,right=2.5cm,top=2.5cm,bottom=3cm]{geometry}
\fontdimen2\font=1.2\fontdimen2{\jot}{5pt}

\titleformat{\section}{\large\bfseries}{\thesection.}{4pt}{}
\titlespacing{\section}{0pt}{20pt}{6pt}

\titleformat{\subsection}{\normalfont\bfseries}{\thesubsection.}{4pt}{}
\titlespacing{\subsection}{0pt}{15pt}{6pt}

\titleformat{\subsubsection}{\normalfont\itshape}{\thesubsubsection.}{4pt}{}
\titlespacing{\subsubsection}{0pt}{15pt}{6pt}

\titleformat{\paragraph}{\normalfont\itshape}{\theparagraph.}{4pt}{}
\titlespacing{\paragraph}{0pt}{15pt}{6pt}

\newcommand{\bea}{\begin{eqnarray}}
\newcommand{\eea}{\end{eqnarray}}
\newcommand{\beq}{\begin{equation}}
\newcommand{\eeq}{\end{equation}}

\def\<{\langle}
\def\>{\rangle}

\def\nn{\nonumber}

\def\cO {{\mathcal O}}

\DeclareFontShape{OT1}{cmr}{mx}{n}%
{<->cmr10}{}
\newcommand{\mytitlefont}{\fontseries{mx}\selectfont}
\DeclareMathAlphabet{\titlemath}{OT1}{cmr}{mx}{n}

\begin{document}

\begin{titlepage}

\begin{center}
			
~\\[0.7cm]
			
{\fontsize{20pt}{0pt} \mytitlefont Bootstrap Cone of the Multicritical Deconfined Quantum Critical Point}
			
~\\[0.4cm]
 
Zhijin Li  and Tinhong Shen

~\\[0.1cm]

{\it School of Physics, Southeast University, Nanjing, China, 210096.

Shing-Tung Yau Center, Southeast University, Nanjing, China, 210096.
}~\\[0.2cm]

\end{center}

\vskip0.5cm

The deconfined quantum critical point (DQCP) provides a prominent example of the unconventional phase transitions beyond the Landau-Ginzburg-Wilson paradigm and its nature has been controversial for decades.
The DQCP has been extensively studied and the results lead to two opposite scenarios with pseudo-criticality or multicriticality.
The pseudo-criticality is a prevailing scenario of DQCP which interprets the approximately scale invariant numerical results with the walking behavior near complex fixed points. In contrast, the multicriticality scenario conjectures the DQCP is a unitary fixed point with a relevant $SO(5)$ singlet scalar. In this work we provide substantial evidence for the multicriticality scenario using conformal bootstrap. 
We start with the observation that the large scale Quantum Monte Carlo (QMC) results nearly saturate the bootstrap bounds. After imposing suitable sparseness condition the bootstrap bound forms a sharp cone in the three-dimensional parameter space. The bootstrap cone is close to the QMC data.
We use the navigator algorithm to locate the apex of the cone and extract the extremal solutions. We find striking consistencies between the bootstrap solutions and the fuzzy sphere data of DQCP, including the coefficients in the operator product expansions (OPEs) and the higher spectrum! 
The bootstrap cone unifies the QMC and the fuzzy sphere data into a unitary conformal field theory with a relevant $SO(5)$ singlet scalar, thus strongly supporting the multicriticality scenario of DQCP.
The agreement between the conformal bootstrap, QMC and fuzzy sphere results is a surprise towards solving DQCP and decoding the profound phase diagram of the two-dimensional quantum magnets. 




\end{titlepage}



\section{Introduction}
The two-dimensional quantum magnets on a lattice serves as invaluable models for exploring exotic states of matters and studying the strongly coupled quantum field theories. The system can realize the N\'eel antiferromagnet (AFM), the Valence-Bond-Solid phase (VBS) and the quantum spin liquid (QSL). They are center to high-$T_c$ superconductivity, topological quantum matter, etc. They also provide a bridge among real materials, quantum criticality and the strongly coupled conformal field theories (CFTs). In particular, the AFM-VBS phase transition represents a prominent violation of the classical Landau-Ginzburg-Wilson paradigm known as the deconfined quantum critical point (DQCP) \cite{senthil2004science,senthil2004prb,senthil2023}.

The DQCP describes a direct AFM-VBS quantum phase transition for a system of $SU(2)$ spins on a square lattice \cite{senthil2004science,senthil2004prb,senthil2023}. The AFM phase breaks the spin rotation symmetry $SO(3)_s=SU(2)_s/\mathbb{Z}_2$ and the VBS phase breaks the lattice symmetries $C_4$. A continuous phase transition is prohibited according to the  Landau symmetry breaking theory. However, the DQCP suggests a continuous AFM-VBS  phase transition can be generated by the fractionalized spinons and the emergent gauge field. The spinons are confined on both sides of the phase transition, while they become deconfined and interact through the emergent gauge field only at the critical point. Its field theory description is given by the non-compact $CP^1$ ($NCCP^1$) model
\begin{equation}
    \mathcal{L}=\sum\limits_{i=1,2}|(\partial_\mu-i A_\mu)z_a|^2+\lambda \left(|z_1|^2+|z_2|^2\right)^2.
\end{equation} 
The theory is $SU(2)_f\times U(1)_t$ symmetric in the UV and an $SO(5)$ symmetry enhancement has been observed in the long distance limit of the DQCP \cite{Nahum2015EmergentSO5}.
Alternatively the phase transition can be formulated in the $SO(5)$ non-linear sigma model (NLSM) with a level-1 Wess-Zumino-Witten (WZW) term \cite{tanaka2005,senthil2006}.
More dual field theory descriptions of DQCP have been proposed which are part of an enlarged 3D duality web \cite{seiberg2016, wang2017, Senthil:2018cru}.

Recently there are exciting developments on realizing the proximate DQCP in real materials \cite{Zayed2017Plaquette,Lee2019ShastrySutherlandDQCP,Guo2020HighPressureThermodynamics,Shi2022ExtremeFieldPressure,cui2023,guo2025,Lin2024TwoPlaquetteSO5}. The compound SrCu$_2$(BO$_3$)$_2$ admits the Shastry-Sutherland lattice structure with  plaquette-singlet and the AFM phases. The transition is found to be first order and related to the ``easy-plane" DQCP with an emergent $O(4)$ symmetry \cite{Lee2019ShastrySutherlandDQCP}. Recently it has been proposed that the material admits an extended phase diagram with full plaquette (FP) and empty plaquette (EP) phases, in which the FP-AFM and EP-AFM phase transitions merge at a multicritical point related to the  DQCP with an enhanced $SO(5)$ symmetry \cite{cui2025}. The DQCP plays a central role in the phase diagram.

\subsection{A brief summary of previous studies}
The DQCP has been extensively studied in the past two decades while its nature is still controversial \cite{
Motrunich2004EmergentPhotons,
Sandvik2007EvidenceDQCP,
Melko2008ScalingFan,
Jiang2008FirstOrderAFMVBS,
Kuklov2008GenericFirstOrder,
Sandvik2010ContinuousTransition,
Lou2009SUNHeisenbergVBS,
Banerjee2010ImpurityTexture,
Kaul2012LargeNSUN,
Kaul2012BilayerSUN,
Bartosch2013CorrectionsScaling,
Chen2013DeconfinedFlow,
Pujari2013HoneycombDQCP,Nahum2015EmergentSO5,
Nahum2015ScalingViolations,
Sreejith2015HigherChargeMonopoles,
Sreejith2014CubicDimerMonomer,
Shao2016TwoLengthScales,
Sato2017DiracCompetingOrders,
Liu2019QSHSC,
Zhao2020,
Wang2021SO5NLSM, Liu2022GaplessQSL, Liu2022GaplessQSLGlobal, Zhao:2022dqcpEE, Zhou2024FuzzySphereDQCP, Song2025EvolutionEE, Chen2024SO5NLSMSphere, Deng2024DiagnosingEE, DEmidio2024EntanglementSO5, Chen2024EmergentConformalSO5,Takahashi2024, Zhu:2026Bi, daliao2026}, see \cite{senthil2023} for a recent review. The early quantum Monte Carlo (QMC) studies have found characteristic properties of DQCP, including the emergent symmetries and the deconfined spinon excitations, while no signs of first-order behavior have been observed, indicating a continuous phase transition. Nevertheless, the results show unconventional scaling behavior \cite{Nahum2015ScalingViolations} and the critical exponents are incompatible with the conformal bootstrap bounds \cite{Nakayama:2016jhq,Poland2019,Li:2018lyb}. It has been widely accepted that the phase transitions observed in these simulations are weakly first order.
A key question is whether the phase transitions are dominated by the nearby complex fixed points related to the pseudo-criticality \cite{Nahum2015EmergentSO5, wang2017,Gorbenko2018,Gorbenko2018II,Ma2020} or a nearby multicritical fixed point \cite{Zhao2020,Takahashi2024}. 

In the past few years the pseudo-criticality has been adopted as a prevailing explanation to the weakly first-order phase transition \cite{Nahum2015ScalingViolations,wang2017,Gorbenko2018,Gorbenko2018II,Ma2020}. In this scenario, the renormalization group flow runs extremely slow in a large infrared scale, 
such walking behavior is hard to be distinguished from the real criticality experimentally. In a more comprehensive picture the pseudo-criticality corresponds to a pair of nearby non-unitary fixed points in the complex parameter space. 
Presumably the two complex fixed points are formed through colliding of a pair of stable/unstable fixed points, which is happened at the lower edge of the conformal window. The smoking gun of this mechanism is that the lowest $SO(5)$ singlet scalar crosses the marginality condition from above. There is intriguing evidence for the pseudo-criticality of DQCP. A particularly impressive work is presented in \cite{Zhou2024FuzzySphereDQCP} based on the fuzzy sphere regularization \cite{Zhu:2022gjc,Hu:2023xak}, in which the $SO(5)$ NLSM with a level-1 WZW term is investigated. The approximate conformal symmetry is observed at the phase transition and abundant CFT data  has been estimated. The authors further observed the lowest $SO(5)$ singlet scalar crosses the marginality condition with increasing system size, consistent with the pseudo-criticality. However, the fuzzy sphere result on the scaling dimension of the lowest $SO(5)$ singlet scalar is not well converged yet according to the data available in \cite{Zhou2024FuzzySphereDQCP}, and it becomes more relevant with increasing system size; while the walking behavior requires the lowest singlet scalar being almost marginal.

An alternative scenario for the DQCP has been proposed with a multicritical fixed point \cite{Zhao2020,Takahashi2024}. Different from the pseudo-critical scenario, the multicritical fixed point is unitary and its lowest $SO(5)$ singlet scalar is relevant. The authors observed a weakly first-order phase transition, which is conjectured to be near a multicritical fixed point  that is hard to reach using current QMC implementation \cite{Takahashi2024}.  By allowing a relevant singlet scalar the multicriticality scenario resolves the contradiction between the bootstrap bounds and the QMC results. Furthermore, some of the perturbative critical indices and the numerical results nearly saturate the bootstrap bounds \cite{Li:2018lyb}, and more agreements are shown in \cite{Chester:2023njo}. 
The $SO(5)$ NLSM associated with the $SO(5)$ symmetry breaking interactions has been studied using the fuzzy sphere regularization \cite{Chen2024SO5NLSMSphere,Chen2024EmergentConformalSO5} and the  continuous-field QMC \cite{daliao2026}, in which the DQCP is suggested to be a multicritical fixed point in a more comprehensive phase diagram. However, the CFT data estimated in \cite{Chen2024SO5NLSMSphere} is not consistent to the bootstrap bounds in \cite{Li:2018lyb}.
In the large scale QMC simulation \cite{Takahashi2024}, the low-lying spectrum has been estimated with extremely small statistical errors. A particularly important data in \cite{Takahashi2024} is the scaling dimension of the lowest $SO(5)$
singlet scalar $\Delta_s=2.274(4)$, which is notably smaller than the marginality condition. 

These extraordinary developments lead to a sharp question:
\begin{center}
\fbox{ Is the DQCP related to complex fixed points or a unitary multicritical fixed point?}
\end{center}

In this work, we will present decisive evidence for this enigma. 
We start with the observation that the large scale QMC data is remarkably close to saturate the bootstrap bounds. 
We will employ the conformal bootstrap approach to carve out the parameter space of the presumed DQCP fixed point, which forms a sharp cone and shows striking consistencies with the large scale QMC data and new fuzzy sphere results from different aspects. 
The special advantage of the conformal bootstrap method is that the bootstrap bounds are ensured to be unitary and mathematically strict under given assumptions, thus providing indispensable pivot to decode the nature of DQCP.

\section{Conformal bootstrap and DQCP}
\subsection{Challenges from uncertainties}
The DQCP is a typical strong correlation problem for which the nonperturbative approaches are necessary. The mainstream tools such as the QMC simulation \cite{Takahashi2024} and the fuzzy sphere regularization \cite{Zhou2024FuzzySphereDQCP} have already  uncovered some critical features of the DQCP. However, these approaches face certain uncertainties and a conclusive answer is still not reachable. In particular the QMC simulation \cite{Takahashi2024} has been implemented on a large system which is sufficient to unambiguously identify the tiny discontinuity at the weakly first-order phase transition and provides critical indices with small statistical errors. Nevertheless, the conjectured multicritical fixed point is inaccessible for the current QMC implementation due to the sign problem. The existence of the multicritical fixed point can not be directly verified. The fuzzy sphere approach has observed the approximate conformal symmetry at the transition point; moreover, thanks to the $S^2\times \mathbb{R}$ geometry and the state-operator correspondence, it has generated rich CFT data for the DQCP. Nevertheless, it remains hard to distinguish the exact and approximate conformal symmetries  and part of the low lying spectrum is not well converged yet. As a result, it is not wholly ascertained that the pseudo-criticality scenario is the only explanation of the fuzzy sphere results. In any case, the true answer should be able to explain the QMC and fuzzy sphere results.  

\subsection{Resolution from strict bootstrap bounds}

In conformal bootstrap, the unitarity is unambiguously implemented \cite{Poland2019,Rattazzi:2008pe,Poland:2016chs,Rychkov:2023wsd}. This can help to eliminate the uncertainties in the QMC and fuzzy sphere results. Specifically
we employ conformal bootstrap method to
strictly determine whether the QMC results are indeed near a unitary multicritical fixed point and compare the fuzzy sphere data with the extremal bootstrap solutions.

The numerical conformal bootstrap is based on the conformal block expansion of the conformal four-point correlators
\begin{align}
    \langle \cO_1(x_1)\cO_2(x_2)\cO_3(x_3)\cO_4(x_4)\rangle &=\frac{1}{x_{12}^{\Delta_1+\Delta_2}x_{34}^{\Delta_3+\Delta_4}}\left(\frac{x_{24}}{x_{14}}\right)^{\Delta_{12}}\left(\frac{x_{14}}{x_{13}}\right)^{\Delta_{34}}\sum\limits_\cO\lambda_{12\cO}\lambda_{34\cO}g_{\Delta,\ell}^{\Delta_{12},\Delta_{34}}(u,v), \nn\\
    u&\equiv\frac{x_{12}^2x_{34}^2}{x_{13}^2x_{24}^2},~~v\equiv\frac{x_{14}^2x_{23}^2}{x_{13}^2x_{24}^2},
\end{align}
in which $\cO_i$ are the scalar operators, $\Delta_{ij}\equiv \Delta_i-\Delta_j$, $x_{ij}\equiv |x_i-x_j|$, and $g_{\Delta,\ell}^{\Delta_{12},\Delta_{34}}(u,v) $ gives the conformal block for the spin $\ell$ operator with scaling dimension $\Delta$. The above $s$-channel conformal block expansion can be alternatively expanded in the $t$-channel which gives the crossing equation
\begin{equation}
    v^{\frac{\Delta_2+\Delta_3}{2}}\sum\limits_\cO \lambda_{12\cO}\lambda_{34\cO}g_{\Delta,\ell}^{\Delta_{12},\Delta_{34}}(u,v)=u^{\frac{\Delta_1+\Delta_2}{2}}\sum\limits_\cO \lambda_{14\cO}\lambda_{32\cO}g_{\Delta,\ell}^{\Delta_{14},\Delta_{32}}(v,u). \label{cseq}
\end{equation}
In bootstrap computations, the operators in the crossing equations (\ref{cseq}) are truncated to sufficiently high spins and the semidefinite program \cite{Simmons-Duffin:2015qma,Landry:2019qug} is employed to obtain constraints on the CFT data.
The conformal bootstrap has already been applied to study DQCP in \cite{Nakayama:2016jhq,Li:2018lyb,He:2020azu, Chester:2023njo,Chester:2025uxb}. The bootstrap bounds have excluded the original proposal of the DQCP with no relevant $SO(5)$ singlet scalar. 
On the other hand, in the multicriticality scenario the critical indices corresponding to the $2\pi$- and $4\pi$-monopoles in $CP^1$ model \cite{Metlitski2008, Dyer:2015zha}  nearly saturate the bootstrap bound \cite{Li:2018lyb}. Such coincidence has been  confirmed further for $6\pi$- and $8\pi$-monopoles \cite{Chester:2023njo}.

\begin{figure}
    \centering
 \includegraphics[width=0.45\linewidth]{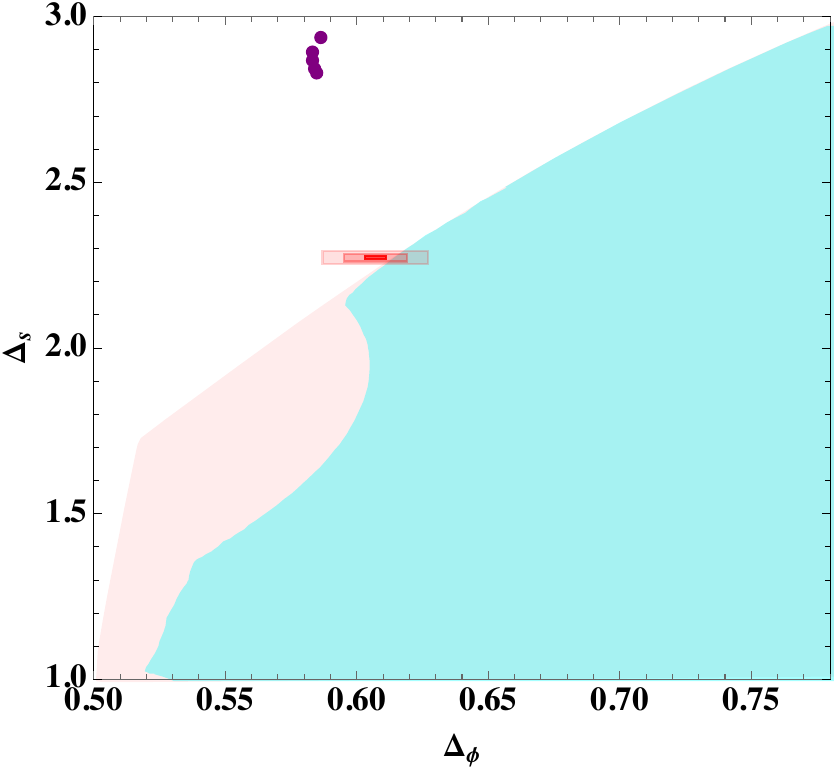}   ~  \includegraphics[width=0.45\linewidth]{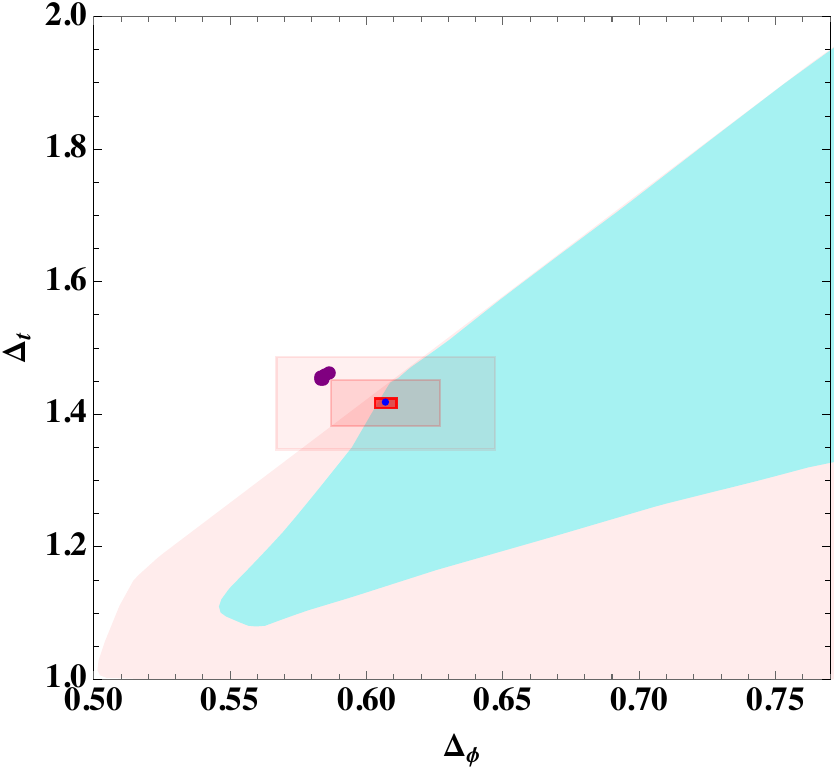}     
    \caption{Bootstrap bounds on the $SO(5)$ singlet scalar $s$ (left) and traceless symmetric scalar $t$ (right). The pink regions denotes the bootstrap allowed regions for general unitary CFTs. The cyan regions are obtained with assumptions $\Delta_{s'}>4.8$ (left) and $\Delta_{t'}>3.95$ (right). The red rectangles represent the QMC results with $1,5$ and $10\sigma$ error bars. The purple dots give the fuzzy sphere results \cite{Zhou2024FuzzySphereDQCP} at different system sizes. }
    \label{fig:2dbd}
\end{figure}

In Figure \ref{fig:2dbd} we show the bootstrap constraints on the new results from the large scale QMC simulation \cite{Takahashi2024}
and fuzzy sphere regularization \cite{Zhou2024FuzzySphereDQCP}. The bounds are obtained by bootstrapping the four-point correlator of the $SO(5)$ vector scalar $\phi_i$: $\langle \phi_i\phi_j\phi_k\phi_l \rangle$. The bootstrap bounds apply to any unitary CFTs with $SO(5)$ global symmetry.
The first notable signature of the bootstrap bounds is the kink which is saturated by the critical $O(5)$ vector model, as firstly observed in the seminal work \cite{Kos:2013tga}. Such saturation of the bootstrap bounds eventually leads to precise numerical solutions of the critical $O(N)$ vector models \cite{Chester:2019ifh,Chester:2020iyt,Chang:2024whu}.
Besides, the large scale QMC results \cite{Takahashi2024}
\begin{equation}
    \Delta_\phi=0.607(4),~~ \Delta_s=2.273(4),~~ \Delta_t=1.417(7) \label{QMC}
\end{equation}
also nearly saturate the bootstrap bounds! Inspired by the bootstrap solutions to the critical $O(5)$ vector models, it is absorbing to conjecture that the QMC data may correspond to a full-fledged unitary CFT which could be numerically solved by conformal bootstrap. The highly nontrivial coincidence between the QMC data and the bootstrap bounds paves the way for bootstrapping the nature of DQCP.


\section{Start of the bootstrap journey}
We start with bootstrapping the critical $O(5)$  vector model, which saturates the general bootstrap bounds on unitary $O(5)$ symmetric CFTs. It is a landmark in the landscape of $SO(5)$ CFTs and provides an instructive example for bootstrapping the DQCP. Moreover, the $O(5)$ vector model represents the universality class of the $SO(5)$ theory of high-$T_c$ superconductivity \cite{Zhang1997SO5,Demler2004}. The $SO(5)$ symmetry is realized at a multicritical fixed point which connects to the two critical lines corresponding to the antiferromagnetic order and the $d$-wave superconducting order. The $SO(5)$ theory of high-$T_c$ superconductivity can be described by the classical Landau-Ginzburg-Wilson theory with $O(2)\oplus O(3)$ symmetry
\begin{equation}
    \mathcal{L}=\frac{1}{2}(\partial_\mu\phi_1)^2+\frac{1}{2}(\partial_\mu\phi_2)^2+\lambda_1 \phi_1^4+\lambda_2 \phi_2^4+u \phi_1^2\phi_2^2,
\end{equation}
where $\phi_{1}/\phi_{2}$ are the $O(2)/O(3)$ vectors. At the multicritical fixed point there is $O(2)\oplus O(3)\Rightarrow O(5)$ symmetry enhancement. The multicritical fixed point is found to be unstable under perturbation of the parameter $u$ \cite{Calabrese2003,Hasenbusch2005}.
Such bicritical phase diagram is reminiscent to the multicriticality scenario of the DQCP proposed in \cite{Takahashi2024}. 

\begin{figure}
    \centering
    \includegraphics[width=0.6\linewidth]{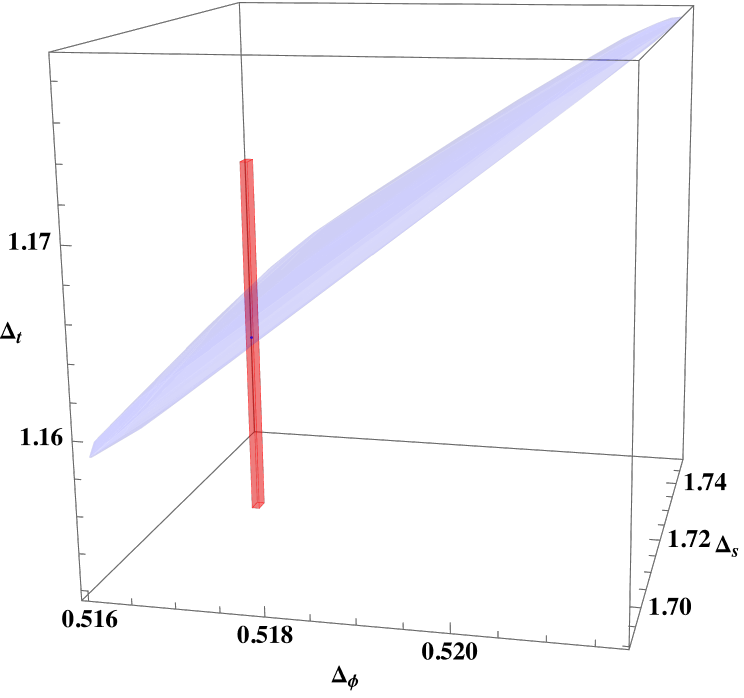}
    \caption{Bootstrap island for the critical $O(5)$ vector model. The red cuboid represents the MC results \cite{Hasenbusch2005,Hasenbusch2022}. }
    \label{fig:O5island}
\end{figure}

The bootstrap bounds of the critical $O(5)$ vector model is shown in Figure \ref{fig:O5island}, in which the allowed parameters are restricted to a small island. The red region shows the Monte Carlo results \cite{Hasenbusch2005,Hasenbusch2022}
\begin{equation}
    \Delta_\phi=0.516985(45), ~~\Delta_s=1.7182(10),~~ \Delta_t=1.162(10),
\end{equation}
in which $\phi,s,t$ denote the lowest scalars in the $SO(5)$ vector, singlet and traceless symmetric representations.
The bound is obtained by bootstrapping the mixed four-point correlators of $\phi_i/s$:  $\langle \phi_i\phi_j\phi_k\phi_l\rangle$, $\langle \phi_i\phi_j ss\rangle$, $\langle \phi_i s\,\phi_j s\rangle$, $\langle ssss\rangle$. Besides, we have introduced the sparseness condition of the spectrum that the $\phi,s,t$ are the only relevant primary scalars in the representations. The island in Figure \ref{fig:O5island} is part of the 3D $O(N)$ archipelago firstly discovered in \cite{Kos:2015mba}. We are particularly interested in its higher spectrum shown in Table \ref{tab:delta-analytic}. The $s'/t'$ denote the primary scalars next to the $s/t$ of the same $SO(5)$ representations.
The bootstrap results are estimated at $\Delta_\phi=0.516985(45), \Delta_s=1.7182(10)$. We use the navigator algorithm \cite{Reehorst2021} to find the maximum $\Delta_t$ and extract the extremal spectrum at the boundary. 
The bootstrap data is well consistent with the perturbative and the MC results \cite{Calabrese2003,Hasenbusch2005,Hasenbusch2022,Calabrese2002,Bonati2025}.
\begin{table}[h]
\centering
\renewcommand{\arraystretch}{1.2}
\begin{tabular}{lllll}
\toprule
\hline
Method & \(\Delta_{t}\) & \(\Delta_{s'}\) & \(\Delta_{{t'}}\) & \(\Delta_{t_{4}}\) \\
\midrule
Bootstrap 
&
\(1.179\)
&
\(3.811\)
&
3.446
&
---
\\
\(\epsilon\)-expansion\cite{Calabrese2003}
&
\(1.168(8)\)
&
\(3.783(26)\)
&
\(3.441(13)\)
&
\(2.802(11)\)
\\
3D expansion \cite{Calabrese2002}
&
\(1.21(5)\)
&
\(3.790(15)\)
&
---
&
\(2.811(10)\)
\\
MC result \cite{Hasenbusch2022}
&
\(1.162(10)\)
&
\(3.754(7)\)
&
---
&
\(2.820(15)\)
\\
\hline
\bottomrule
\end{tabular}
\caption{Scaling dimensions of the scalar operators from different methods. }
\label{tab:delta-analytic}
\end{table}

We will refer to the spectrum in  Table \ref{tab:delta-analytic} in the DQCP bootstrap.
Both the critical $O(5)$ vector model and the QMC results in \cite{Takahashi2024} nearly saturate the bootstrap bounds. These extremal solutions are expected to have sparse spectrum. To carve out the CFT data for DQCP, it requires to introduce the sparseness conditions which are at least strong enough to exclude the extremal solution of the critical $O(5)$ vector model, so one expects $\Delta_{s'}>3.811, \Delta_{t'}>3.446$ for DQCP.


\section{The bootstrap cone of DQCP}
\begin{figure}
    \centering
    \includegraphics[width=0.6\linewidth]{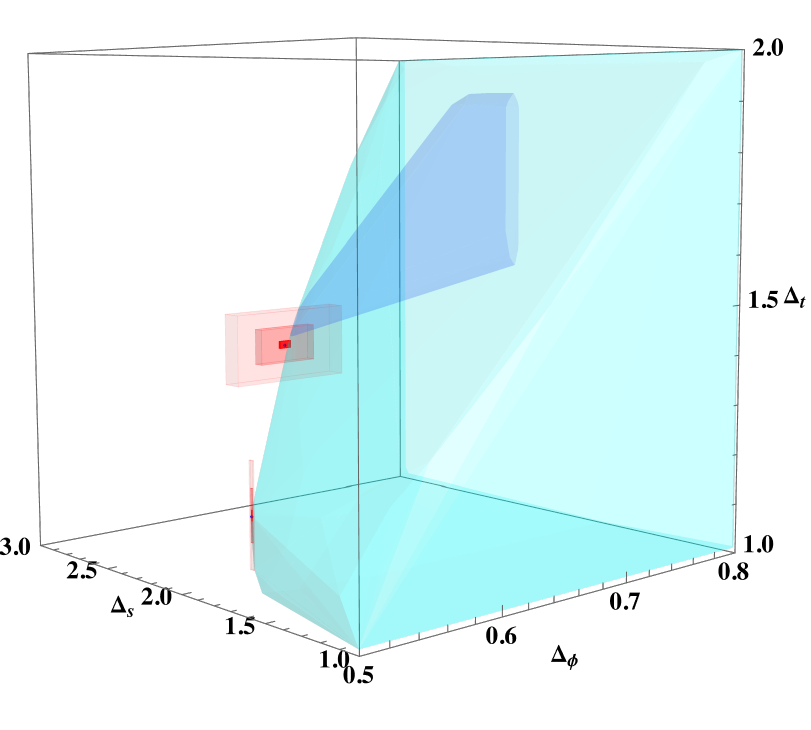}
    \caption{The cyan (blue) region represents the bootstrap bound for $SO(5)$ CFTs (with assumptions $\Delta_{s',t'}>3$).  The two separated red cuboids denote the QMC results for the critical $O(5)$ vector model and DQCP.}
    \label{fig:3dbd}
\end{figure}

We now explore deeper structures in the conformal parameter space beyond the apparent coincidence between the QMC data and the bootstrap bounds. 
The results are obtained by bootstrapping the four-point correlator of the $SO(5)$ vector scalar $\phi_i$: $\langle \phi_i\phi_j\phi_k\phi_l\rangle$. 
In Figure \ref{fig:3dbd} we present the bootstrap bounds in the three dimensional parameter space for the scaling dimensions $(\Delta_\phi, \Delta_s, \Delta_t)$. The cyan region represents the bootstrap allowed parameters for any $SO(5)$ symmetric unitary CFTs. The QMC results for the $O(5)$ vector model \cite{Hasenbusch2005,Hasenbusch2022} and the DQCP \cite{Takahashi2024} are shown in the red cuboids with $1,5$ and $10\sigma$ error bars. Both QMC results nearly saturate the three dimensional bootstrap bound. In particular, the DQCP critical indices (\ref{QMC})
are measured on the weakly first-order phase transition line  which is  conjectured to be near a unitary multicritical fixed point \cite{Takahashi2024}. The bootstrap bound in Figure \ref{fig:3dbd} strongly supports this conjecture: the center of the QMC data is slightly outside of the bootstrap bound, consistent with the fact that it is estimated from the weakly first-order line; besides, it is especially close to the bootstrap bound in which the solutions are unitary. Since the presumed unitary CFT nearly saturates the bootstrap bound, we can extract more information about the theory using conformal bootstrap. 

\begin{figure}
    \centering
    \includegraphics[width=0.65\linewidth]{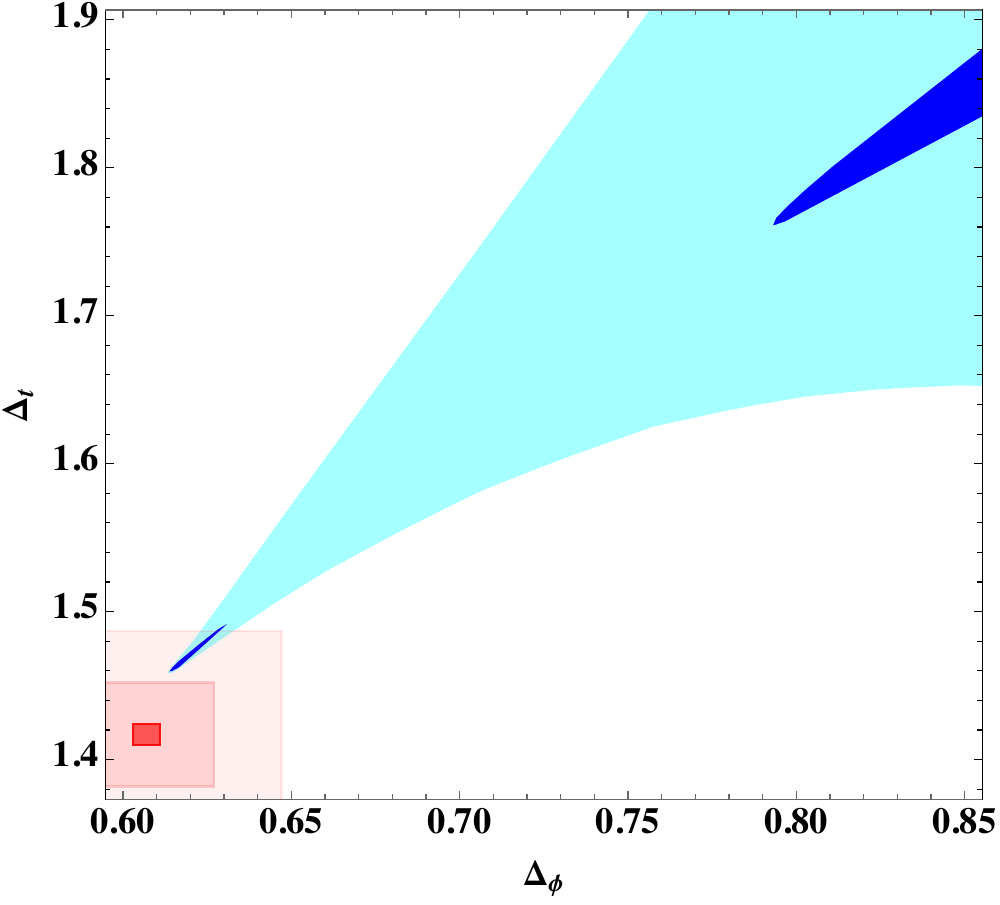}
    \caption{Bootstrap bounds on the plane with fixed $\Delta_s=2.273$. The cyan (blue) region represents the bound with assumptions $\Delta_{s'}>3, \Delta_{t'}>3$ ($\Delta_{s'}>4.5, \Delta_{t'}>3.95$).}
    \label{fig:gap3-45}
\end{figure}

We need to introduce sparseness conditions to carve out the conformal parameter space. 
In the $O(5)$ vector model bootstrap, the gap conditions $\Delta_{s',t'}>3$ have been imposed. The conditions also applies to the DQCP and the related bootstrap bound shrinks into a long strip given by the light blue region in Figure \ref{fig:3dbd}. In fact the presumed DQCP nearly saturates the bootstrap bound in the deeper region and one expects stronger sparseness conditions than those of the critical $O(5)$ vector model in Table \ref{tab:delta-analytic}. In Figure \ref{fig:2dbd} it is shown that by introducing the sparseness condition $\Delta_{s'}>4.8$ ($\Delta_{t'}>3.95$), the bootstrap bound on $\Delta_s$ ($\Delta_t$) forms sharp kinks near the QMC data. This gives the suitable sparseness condition to carve out the extremal bootstrap solution near the QMC data. The fuzzy sphere results suggest $\Delta_{t'}=4.351$ and no conformal primary $SO(5)$ singlet scalar is identified for $\Delta_{s'}<5.5$, notably above the lower gaps we imposed $\Delta_{s'}>4.8,~ \Delta_{t'}>3.95$.

\begin{figure}
     \centering
     \includegraphics[width=0.5\linewidth]{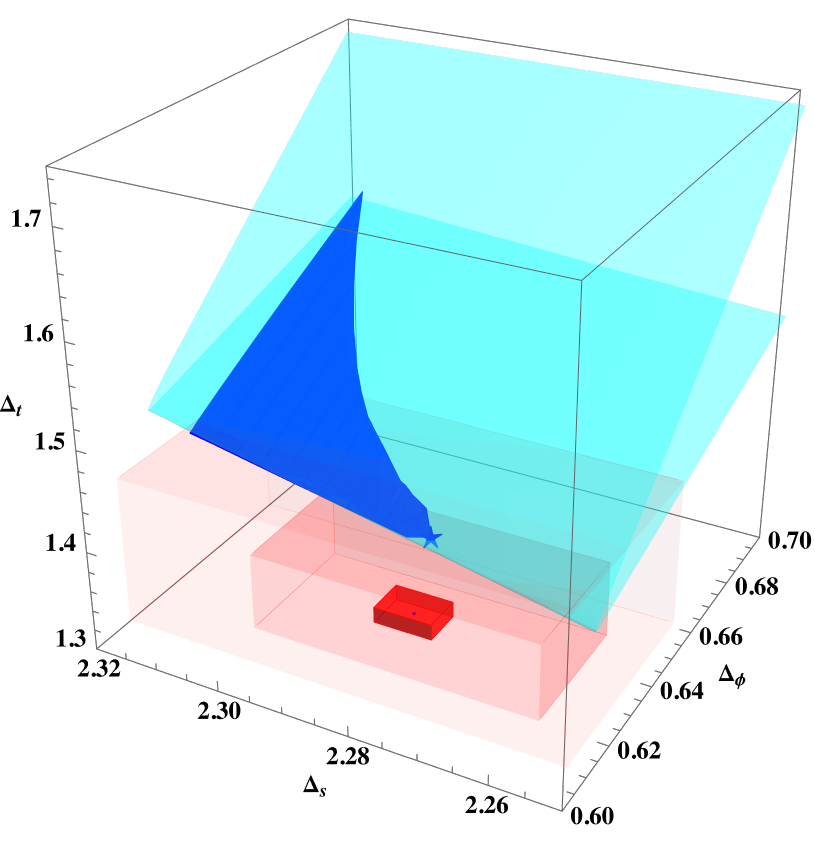}
    \caption{The cyan region gives the bootstrap bound with assumptions $\Delta_{s',t'}>3$. The blue cone shows the bootstrap bound with assumptions $\Delta_{s'}>4.8$ and $\Delta_{t'}>3.95$. The star denotes the apex of the cone.}
    \label{fig:cone}
\end{figure}

We show the bootstrap bounds with different sparseness conditions in Figures \ref{fig:gap3-45} and \ref{fig:cone}. 
The most distinct signature of the new bootstrap bounds is that the tip of the long strips collapses into a sharp cone. On the profiles with fixed $\Delta_s$, the previous long strips break off and form closed islands on the tip, see Figure \ref{fig:gap3-45} for the bootstrap bounds on the profile $\Delta_s=2.273$. The island extends with larger $\Delta_s$ and forms a sharp cone shown in Figure \ref{fig:cone}.
We use the navigator algorithm to find the apex of the cone, which is marked by a star. It is remarkable that even though the whole bootstrap bound has shrunk significantly after introducing the sparseness conditions, the bootstrap cone is rather stable under these constraints and adjacent to the large scale QMC result, see Appendix \ref{appdxB} for more results. The apex of the cone gives a representative solution to the DQCP. It corresponds to a unitary CFT with a relevant $SO(5)$ singlet scalar, that is to say, a multicritical fixed point!

\subsection*{From cone to island}
\begin{figure}
    \centering
    \includegraphics[width=0.475\linewidth]{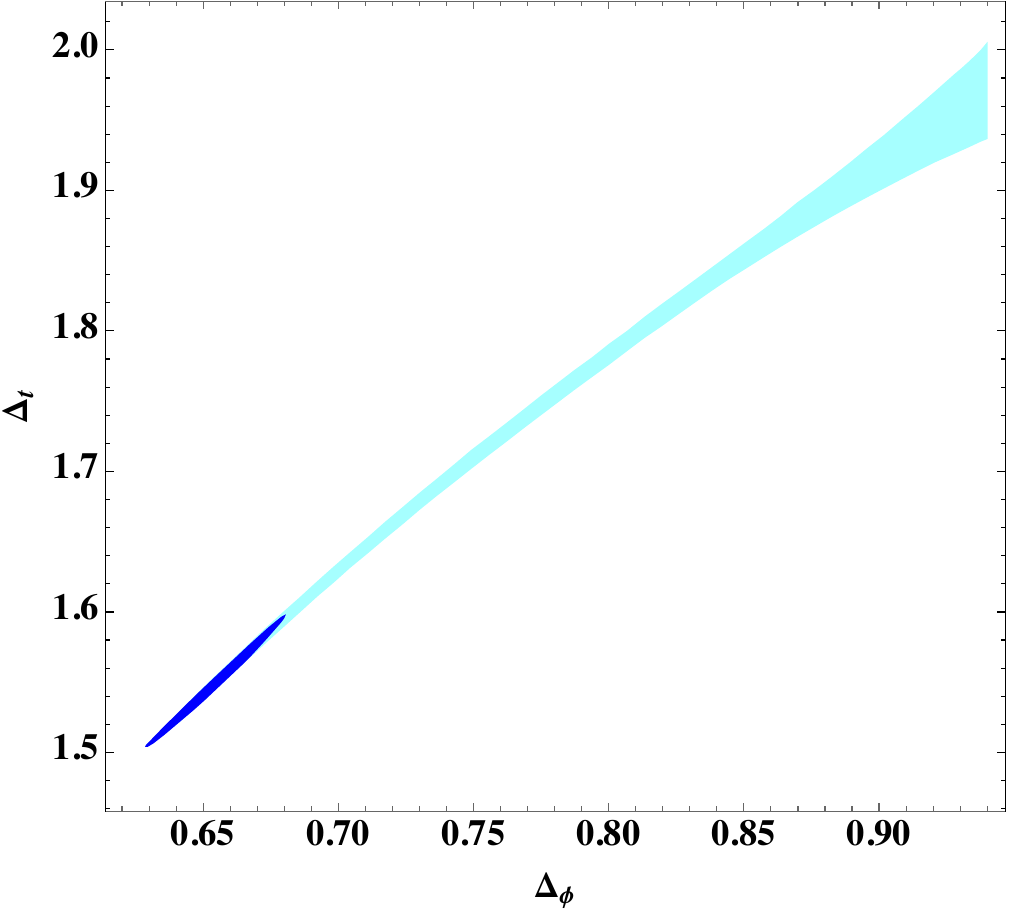} ~ \includegraphics[width=0.49\linewidth]{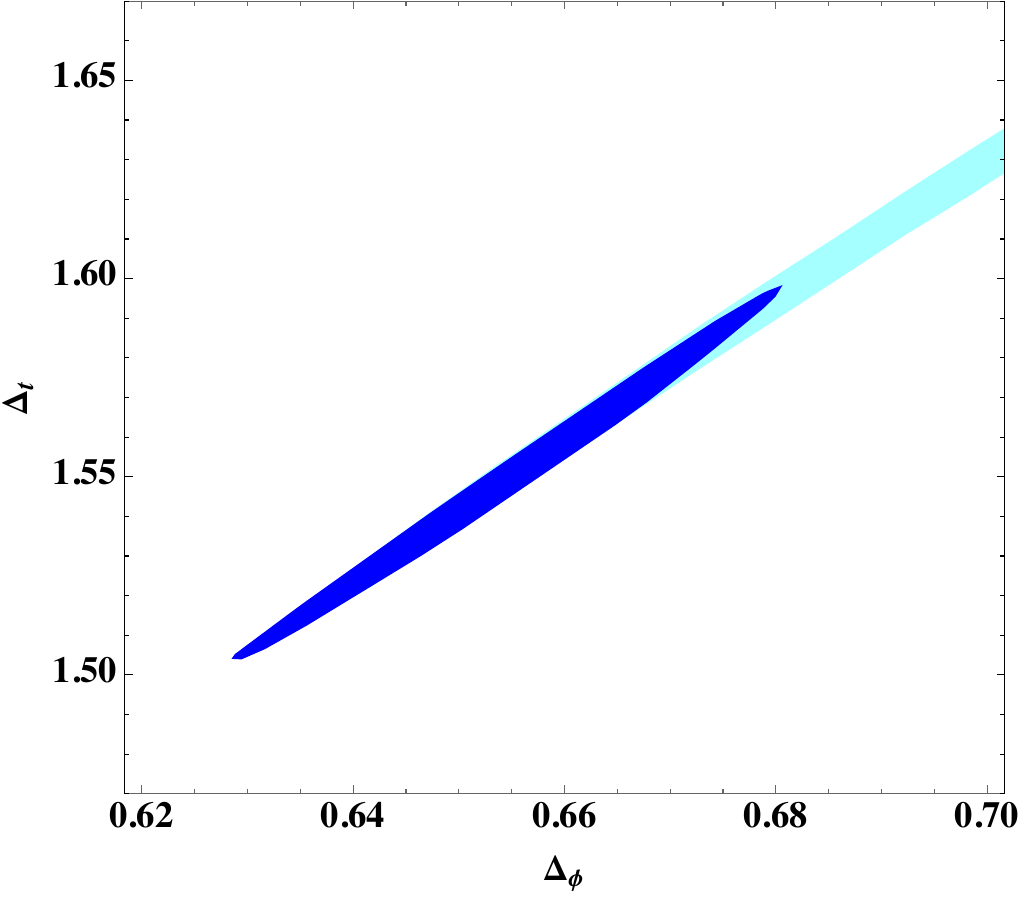}
    \caption{The cyan region gives the bootstrap bound on the slice $\Delta_{s'}=2.35$ with assumptions $\Delta_{s'}>4.8$ and $\Delta_{t'}>3.95$. The blue island is obtained by requiring the locality of the bootstrap solutions.}
    \label{fig:stripisland}
\end{figure}
A fascinating goal is to isolate the bootstrap cone into an island to obtain precise solution of DQCP, like the miracle achievements for the 3D critical Ising and $O(N)$ vector models \cite{Kos:2015mba,Kos:2014bka}. The bootstrap cone is connected to the continent through long strips, see Figure \ref{fig:stripisland} for an example, which is obtained on the profile $\Delta_s=2.35$ with conditions $\Delta_{s'}>4.8$, $\Delta_{t'}>3.95$.
We find that the extremal spectrum on the long strip has no conserved stress tensor $T_{\mu\nu}$, suggesting the solutions are non-local theories. By introducing a lower bound on the OPE coefficient $\lambda_{T}$ in the bootstrap setup, we can rule out such non-local theories and the new bootstrap bound on this profile becomes an island. The island continues with increasing $\Delta_{s}$ and eventually merges with the continent, so there are unknown local theories connected the bootstrap cone and the continent. To generate a bootstrap island, more sophisticated bootstrap setup is needed to classify and rule out these solutions.
We leave a careful exploration along this direction in the next work.

\section{Bootstrap cone meets fuzzy sphere}
We show that the apex of the bootstrap cone is indeed corresponding to the DQCP. Using the extremal functional method we can extract the higher spectrum and also the OPE coefficients in this particular bootstrap solution. A large set of DQCP CFT data is generated in an excellent fuzzy sphere work \cite{Zhou2024FuzzySphereDQCP}. The fuzzy sphere system only realizes the conformal symmetry approximately so the unitarity of the data is undetermined yet. As a result of the $S^2\times \mathbb{R}$ geometry and the state-operator correspondence, it can classify the 
CFT data of DQCP systematically. This makes a careful comparison between the bootstrap solution and the DQCP data possible. 

We use the navigator algorithm \cite{Reehorst2021} to locate the apex of the cone, from which the extremal solution is extracted.  Details on the bootstrap cones and apexes are presented in Appendix \ref{appdxB}. Some fundamental OPE coefficients and the low-lying spectrum are shown in Tables \ref{tab:ope} and \ref{tab:spectrum}. 
It is striking that the bootstrap results on the OPE coefficients of the conserved current $J_\mu$ and the scalar $t$ are all remarkably close to the fuzzy sphere data. For the OPE coefficient of the lowest $SO(5)$ singlet scalar $s$, there is mild difference in accord with that the fuzzy sphere result on the singlet scalar is not well converged yet. Furthermore, the spectra from conformal bootstrap and fuzzy sphere are also highly consistent with each other except mild differences in the singlet sector. In \cite{Zhou2024FuzzySphereDQCP} the scaling dimension $\Delta_s$ has been estimated at different system sizes, which decreases monotonously with increasing system size. A naive extrapolation predicts a much smaller $\Delta_s$, indicating the discrepancy could be remedied with well converged fuzzy sphere data on $\Delta_s$.  Overall the impressive agreement between the bootstrap and fuzzy sphere results clearly shows that the bootstrap cone is closely related to the DQCP.

\begin{table}[h]
\centering
\renewcommand{\arraystretch}{1.2}
\begin{tabular}{lllll}
\toprule
\hline
Method & \(\lambda_{\phi\phi J}\) & \(\lambda_{\phi\phi T}\) & \(\lambda_{\phi\phi t}\) & \(\lambda_{\phi\phi s}\) \\
\midrule
Fuzzy sphere ~~
&
\(0.771(3)\)~~
&
$0.348(12)$\footnotemark[1]~~
&
1.242(7)~~
&
0.235(8)
\\
Bootstrap
&
\(0.765\)
&
\(0.351\)
&
\(1.260\)
&
\(0.320\)
\\
\hline
\bottomrule
\end{tabular}
\caption{OPE coefficients from the fuzzy sphere and the apex of the bootstrap cone.}
\label{tab:ope}
\end{table}
\footnotetext[1]{Note in Table \uppercase\expandafter{\romannumeral 3} of \cite{Zhou2024FuzzySphereDQCP}, $\lambda_{\phi\phi T}=0.121(8)$. But this  value is too small to produce the central charge $C_T\approx 6.561$ provided in the paper. Instead, its square root can give an approximate central charge $C_T\approx 6.364$ thus is cited here, while the true fuzzy sphere value needs to be confirmed by the authors.    }

\begin{table}[h]
\centering
\renewcommand{\arraystretch}{1.2}
\begin{tabular}{lllllllllll}
\toprule
\hline
Rep. & $\mathbf{1}$  &  $\mathbf{1}$ & $\mathbf{10}$  &  $\mathbf{10}$ & $\mathbf{10}$  & $\mathbf{10}$ & $\mathbf{10}$  & $\mathbf{15}$ & $\mathbf{15}$ & $\mathbf{15}$   \\
\midrule[0.5pt] 
Spin & $0$ & $2$ & 1 & 1 & 1 & 3 & 3 & 0 & 0 & 2 \\
\midrule[0.5pt] 
F.S.
&
2.831
&
3.000
&
2.000
&
3.164
&
$4.515^*$
&
4.215 &
$5.189^*$
&
1.458 &
4.351
&
3.333
\\
C.B.
&
2.267
&
3.000
&
2.000 
&
2.754 & 4.409 & 4.188 & 5.121  & 1.473 & $4.000^*$ & 3.240 
\\
\hline
\bottomrule
\end{tabular}
\caption{Spectra from the fuzzy sphere and the apex of the bootstrap cone. Following \cite{Zhou2024FuzzySphereDQCP}, here we focus on the operators with scaling dimension $\Delta<5.5$ and spin $\ell<4$. In the bootstrap spectrum, we only consider the stable operators with OPE coefficients $\lambda_{\phi\phi\cO}^2\geqslant 10^{-1}$. Note the fuzzy sphere spectra at $\Delta=4.515, \,5.189$ are listed in \cite{Zhou2024FuzzySphereDQCP} but the authors did not identify them as primary or descendant. In the bootstrap spectrum the operator $\Delta=4.000, \ell=0$ in the $\mathbf{15}$ representation is generated by the gap assumption.  }
\label{tab:spectrum}
\end{table}

The OPE coefficients in Table \ref{tab:ope} suggest $\lambda_{\phi\phi s}\ll \lambda_{\phi\phi t}$. Actually for a large $N$ theory the OPE coefficient in the singlet sector is $1/N$ suppressed so such a relation is expected for the $SO(5)$ DQCP. In the QMC simulation, the lattice operators are found to have small overlap with the $SO(5)$ singlet \cite{Takahashi2024}, which can help to explain the fact that the weakly first-order phase transition appears in a broad scale range  although the lowest singlet scalar is relevant $\Delta_s\simeq 2.273(4)$. In general the overlap between the lattice operators and the singlet scalar $\cO$ depend on both the lattice Hamiltonian and the OPE coefficient $\lambda_\cO$ at the infrared fixed point. A small OPE coefficient $\lambda_{\phi\phi s}$ can partially explain the small overlap between the lattice operators and the singlet scalars. We expect the QMC observations \cite{Takahashi2024} can be quantitatively explained with more bootstrap data.

Now we have completed the jigsaw puzzle: the large scale QMC simulation  generates low-lying critical indices with remarkably small error bars \cite{Takahashi2024}, and the fuzzy sphere method provides systematical CFT data for DQCP \cite{Zhou2024FuzzySphereDQCP}, while the unitarity is still a unresolved problem for both approaches; conformal bootstrap unifies the QMC and fuzzy sphere results into a sharp bootstrap cone, which provides a strict check of the unitarity and other consistency conditions. Combining the three approaches the DQCP is strongly suggested to be a unitary multicritical fixed point. 


\section{Discussion}
Our bootstrap computations have carved out a sharp cone in the three dimensional conformal parameter space.
The apex of the cone is close to the large scale QMC results; moreover, its higher spectrum and the OPE coefficients are highly consistent with the new fuzzy sphere results. This suggests the bootstrap cone is closely related to the DQCP, and it uncovers the nature of DQCP from two aspects: the theory is unitary and its lowest $SO(5)$ singlet scalar is obviously relevant instead of marginal. Thus the pseudo-criticality scenario of the DQCP is disfavored. Our results strongly support that the DQCP is a unitary multicritical fixed point. Moreover, the sharp bootstrap cone paves the way to numerically solve the DQCP with conformal bootstrap.

It is important to develop the DQCP studies using both conformal bootstrap, QMC and fuzzy sphere approaches. 
We expect that by bootstrapping the mixed correlators with other low-lying operators in DQCP, it can generate more precise and comprehensive CFT data and it is tempting to promote the bootstrap cone into an island. The QMC results \cite{Takahashi2024} play important roles in our bootstrap study, it would be valuable to get more QMC data and verify our bootstrap results. The fuzzy sphere data \cite{Zhou2024FuzzySphereDQCP} on the $SO(5)$ singlet sector is not well converged yet and show mild difference with the bootstrap and QMC results. Note the $SO(5)$ singlet $s$ which is relevant in the infrared is not tuned away but treated as a running direction in \cite{Zhou2024FuzzySphereDQCP}. An extra tuning for $s$ is needed to accurately reach the multicritical fixed point, which could generate more precise data and eliminate the gap in the $SO(5)$ singlet sector.
Hopefully the results from conformal bootstrap, QMC and fuzzy sphere will merge at a complete solution of DQCP in the near future.

The highly nontrivial consistency between the bootstrap solution and the fuzzy sphere results is encouraging for bootstrapping conformal gauge theories. It has been a long endeavor for the bootstrap community to generalize the miracle successes of the conformal bootstrap to strongly coupled gauge theories while the results are limited  \cite{Poland2019, Rychkov:2023wsd}.  
Presumably the reasons include that the gauge theories are associated with more complicated dynamics than the Ising-like CFTs and also they locate in the deeper regions in the bootstrap bounds where the CFTs are more crowded and hard to isolate. Another subtle reason is that for the theories with non-$SO(N)$ global symmetries, the crossing equations have degenerated $O(N)$ symmetric positive structures \cite{Li:2020bnb,Li:2020tsm}, which make it hard to generate bootstrap solutions with proper global symmetries. For instance, the bootstrap study of $N_f=4$ QED$_3$ has led to strong constraints on the CFT data  while it remains challenging to reproduce the proper $SU(4)_f\times U(1)_t$ symmetric spectrum \cite{Albayrak:2021xtd}.  In contrast, our bootstrap results successfully generate the higher spectrum and OPE coefficients of the $SO(5)$ DQCP obtained from fuzzy sphere method. This provides an instructive example for bootstrapping comprehensive CFT data of conformal gauge theories. 

The $O(4)$ symmetric DQCP has been widely accepted to be first order, see \cite{Chester:2024waw,dumitrescu2024symmetrybreakingmonopolecondensation,Yang:2026SO5O4Flow,zou2025O4} for recent studies. Inspired by the multicriticality of its $SO(5)$ counterpart, one may wonder whether the $O(4)$ DQCP is also multicritical. 
Some QMC simulations of the $O(4)$ DQCP have observed clear discontinuity instead of the pseudo-critical behavior \cite{Serna2019ApproximateO4,ZhaoWeinbergSandvik2019O4}. Continuous phase transitions were observed in certain numerical simulations \cite{Qin:2017xqg, Liu2024EmergentSymmetry} while the critical indices are either excluded by the bootstrap bounds or do not saturate the bootstrap bound even with the multicriticality assumption \cite{Li:2021emd}.  The scenario of $O(4)$ DQCP should be significantly different  from this study.

The most interesting application of our results is the realization of the multicritical DQCP in real materials. Recent experimental studies exhibit evidence for realizing the $O(4)$ DQCP in the compound SrCu$_2$(BO$_3$)$_2$ \cite{Zayed2017Plaquette,Lee2019ShastrySutherlandDQCP,Guo2020HighPressureThermodynamics,Shi2022ExtremeFieldPressure,cui2023,guo2025,Lin2024TwoPlaquetteSO5}, whose crystal structure is the closest to the Shastry-Sutherland model \cite{ShastrySutherland1981}. The phase transition is found to be first order, consistent with the previous QMC and bootstrap results on the $O(4)$  DQCP \cite{Guo2020HighPressureThermodynamics,guo2025}. There is a new proposal that the $SO(5)$ DQCP could appear in the phase diagram of SrCu$_2$(BO$_3$)$_2$ by double tuning the hydrostatic pressure and magnetic field \cite{cui2025}. It would be extremely interesting to verify the multicritical bootstrap solution experimentally.

It is hard to verify the multicritical DQCP using QMC simulation directly. 
An extra interaction with a negative coefficient is needed to reach the multicritical fixed point, which leads to the sign problem \cite{Takahashi2024}. 
Instead, one may apply the QMC simulation to study the long-range version of the DQCP, which shares the same conformal window as the short-range DQCP \cite{li2024}. Given  the long-range multicritical DQCP could be realized with the sign-free Hamiltonian, then that would indirectly confirm the existence of the short-range multicritical DQCP. The long-range DQCP has drawn interests recently \cite{YangYaoSandvik2020LongRangeDQCP,Jian2021NonlocalNeelVBS,Xu2022NonlocalDQCP,Romen2024LongRangeAnisotropicHeisenberg}, and it would be insightful to search the multicritical fixed point in the $3$D long-range models.

The continuum field description of DQCP has inspired a web of dualities in 3D quantum field theories  \cite{seiberg2016, wang2017, Senthil:2018cru}, which provide a laboratory to study theoretical properties like the 3D bosonization, dualities and symmetry enhancement. Initially these studies were motivated by the criticality of DQCP and have been receded following the pseudo-criticality interpretation \cite{Nahum2015ScalingViolations,Gorbenko2018,Gorbenko2018II,Ma2020}, which now need to be modified accordingly. The $NCCP^{N-1}$ model has a critical number $N_c$ below which the infrared unitary fixed point disappears.  The critical flavor number is estimated from the anomalous logarithmic behavior of the entanglement entropy $N_c\in (7, 8)$ \cite{Song:2023wlg}. The multicriticality of the DQCP suggests the evolution of the $NCCP^{N-1}$ fixed point is more complicated than expected before.

The DQCP plays a central role in classifying the quantum phases of matter, thus a solid understanding on the nature of the DQCP can provide a foundation to explore the phase diagram, including the line of the weakly first-order phase transition \cite{Takahashi2024}, the phase transitions among the N\'eel, VBS and the quantum spin liquid phases \cite{Chen2024SO5NLSMSphere,Chen2024EmergentConformalSO5,daliao2026}. Conformal bootstrap results are expected to expedite studies along this direction.

\section*{Acknowledgements}
We are grateful to Meng Cheng, David Poland, Hui Shao, Ning Su, Cenke Xu and Wei Zhu for helpful discussions and communications at different stages of this work. ZL would like to thank the organizers of the conference {``Frontier Forum on Exactly Solvable Models in Statistical Physics and Quantum Theories"} at Thousand-island Lake for the support during the last stage of this work.
This research was supported by the Startup Funding  4007022314 of the Southeast University, and the National Natural Science Foundation of China funding No. 12375061.

\appendix

\section{Normalization of the OPE coefficients}
We explain the normalization for the OPE coefficients $\lambda_\cO$ in our bootstrap computations:
\begin{equation}
    \phi_i\times \phi_j \sim \lambda_{\cO_S}\, \cO_S+\lambda_{\cO_T}\, \cO_T+\lambda_{\cO_A}\, \cO_A.
\end{equation}
In general the OPE coefficients $\lambda_\cO$ are affected by both the normalizations of the operator $\cO$ itself and also the bootstrap crossing equations from the four-point correlator $\langle\phi_j\phi_j\phi_k\phi_l\rangle$. 

We use the navigator algorithm to detect the boundary of the bootstrap bound, and extract the spectrum and OPE coefficients $\lambda'_\cO$ in the extremal solutions. The OPE coefficients $\lambda'_\cO$ depend on the normalization factors. Then we use the same bootstrap implementation to extract the extremal solutions with $\lambda'_{\textrm{f.s.}}$ close to the unitary bound $\Delta_\phi=\frac{1}{2}$. The solutions are approximately given by the theory with $N$ free scalars $\lambda_{\textrm{f.s.}}$. The normalization factors can be unambiguously fixed from  $\lambda'_{\textrm{f.s.}}/\lambda_{\textrm{f.s.}}$, from which we can determine the OPE coefficient $\lambda_\cO$.

The raw data of the OPE coefficients  is
\begin{equation}
    \lambda'_{\phi\phi J}=11.71,~\lambda'_{\phi\phi T}=0.6162,~\lambda'_{\phi\phi t}=22.07,~\lambda'_{\phi\phi s}=0.5128,
\end{equation}
which leads to the OPE coefficients in Table \ref{tab:ope}. Note for the $SO(5)$ singlet scalar, we adopt the normalization of the scalar $s$
\begin{equation}
   \langle s(x)s(0)\rangle=\frac{1}{x^{2\Delta_s}},~~~~ \langle \phi_i\phi_j s\rangle \propto\lambda_{\phi\phi s}^{\textrm{free}}=\sqrt{\frac{2}{N}}. 
\end{equation}
For the $SO(5)$ traceless symmetric scalar $t$, we take the normalization so that 
\begin{equation}
  \langle t_{ij}t_{kl}\rangle\propto \frac{1}{2}\left(\delta_{ik}\delta_{jl}+\delta_{il}\delta_{jk}-\frac{2}{N}\delta_{ij}\delta_{kl}\right),~~~  \lambda_{\phi\phi t}^{\textrm{free}}=\sqrt{2}
\end{equation}
in the $SO(5)$ free scalar theory. The $SO(5)$ conserved current $J_\mu$ is normalized with
\begin{equation}
    \langle J^{ij}J^{kl}\rangle\propto (\delta_{ik}\delta_{jl}-\delta_{il}\delta_{jk}), ~~~ \lambda_{\phi\phi J}^{\textrm{free}}=\frac{1}{\sqrt{2}}
\end{equation}
in the $SO(5)$ free scalar theory.

\section{The bootstrap cone with different sparseness conditions} \label{appdxB}
\begin{figure}
    \centering
    \begin{tabular}{cc}
    \includegraphics[width=0.48\linewidth]{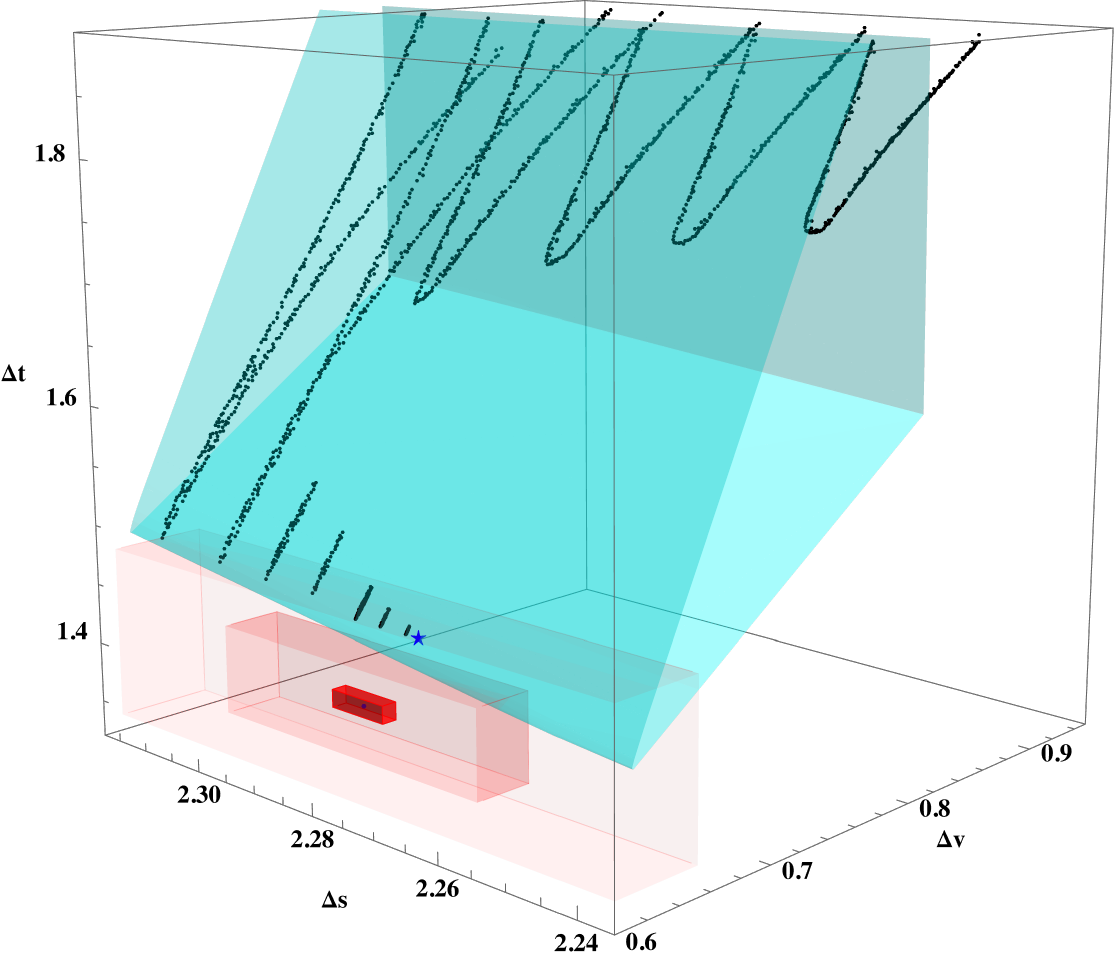}  &\includegraphics[width=0.48\linewidth]{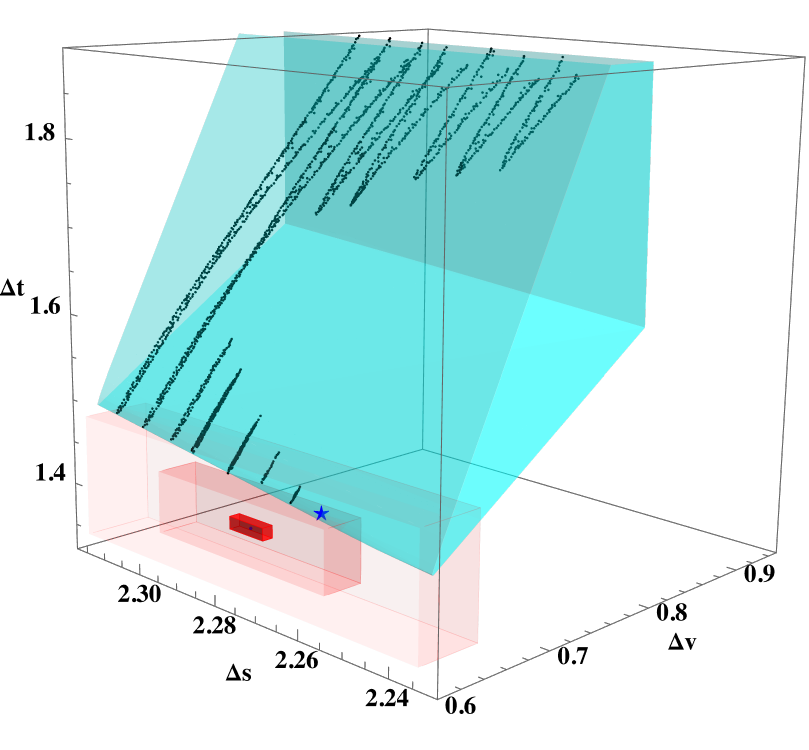}
     \end{tabular}
    \caption{The black dots mark the bootstrap bounds on fixed $\Delta_s$ profiles with assumptions $\Delta_{s'}>4.0, \Delta_{t'}>4.0$ (left) and $\Delta_{s'}>4.5, \Delta_{t'}>3.95$ (right).} 
    \label{fig:moregaps}
\end{figure}

In Figures \ref{fig:cone} and \ref{fig:moregaps} we show that the bootstrap cone locates stably near the QMC results with different assumptions on the spectrum $\Delta_{s',t'}$. Here we present the detailed information on the position of the apexes of the bootstrap cones and their corresponding CFT data. 

In Table \ref{tab:dgaps}  we show the low-lying spectrum of the extremal solutions with different assumptions. Both the spectrum and the OPE coefficients are stable with the assumptions listed in the table. Note that with $\Delta_{s'}>4.5, 4.8$ the lowest spin $1$ operator in the $SO(5)$ antisymmetric representation is slightly above the unitary bound. We can require the existence of the $SO(5)$ conserved current and introduce a small gap $\Delta_{J'}>2.2$ for the next spin 1 operator in the bootstrap implementation. Then the navigator algorithm can find an extremal solution with a conserved current, and the new apex is just slightly shifted from the original position at the order $O(10^{-4})$. The extra spectrum is almost the same as before. Nevertheless, the OPE coefficient $\lambda_{\phi\phi J}$ is affected by the  fake operator at the imposed gap $\Delta_{J'}=2.2$, namely the sharing effect \cite{Liu:2020tpf}. In the main body of this work we adopted the CFT data with gaps $\Delta_{s',t'}>4.0$, which is not affected by the sharing effect and also close to the QMC data.

\begin{table}[h]
\centering
\renewcommand{\arraystretch}{1.2}
\begin{tabular}{c|lllllllll}
\toprule
\hline
~Gaps $(s', t')$  & ~$\Delta_J$ & \(\lambda_{\phi\phi J}\) & $\Delta_T$ &\(\lambda_{\phi\phi T}\) & $\Delta_t$ &\(\lambda_{\phi\phi t}\) & $\Delta_s$ &\(\lambda_{\phi\phi s}\) & $\Delta_\phi$ \\
\midrule
$(4.0, 3.95)$
&  ~2.000  &
\(0.758\)
&  3.000 &
$0.346$
&  $1.453$  &
$1.261$
&  $2.234$  &
$0.323$ & $0.612$
\\
$(4.2, 3.95)$
& ~2.000  &
\(0.757\)
&  3.000 &
$0.345$
&  $1.453$  &
$1.261$
&  $2.240$  &
$0.324$ & $0.610$
\\
$(4.5, 3.95)$
& ~2.015  &
\(0.765\)
&  3.000 &
$0.345$
&  $1.448$  &
$1.261$
&  $2.258$  &
$0.323$ & $0.612$
\\
$(4.8, 3.95)$
& ~2.049  &
\(0.788\)
&  3.000 &
$0.343$
&  $1.462$  &
$1.263$
&  $2.274$  &
$0.327$ & $0.616$
\\
$(4.0,4.00)$
&  ~$2.000$   &
\(0.765\)
&   $3.000$  &
\(0.351\)
&   $1.473$  &
\(1.260\)
&   $2.267$  &
\(0.320\) & $0.617$
\\ 
\hline
\bottomrule
\end{tabular}
\caption{Low-lying operators and their OPE coefficients at the apexes of the bootstrap cones with different assumptions. }
\label{tab:dgaps}
\end{table}

\section{Operators with small OPE coefficients}
The bootstrap spectrum shown in Table \ref{tab:spectrum} is selected with $\Delta<5, ~\ell\leqslant3$ and the OPE coefficients $\lambda_{\phi\phi\cO}^2\geqslant 10^{-1}$.
In the bootstrap spectrum there are three extra operators with smaller OPE coefficients. Higher spectrum with small OPE coefficients in the bootstrap extremal solutions can be affected by the numerical effects and not all of them are physical. Here we show their comparisons with the fuzzy sphere results. 

In Table \ref{tab:mismatch} we show the bootstrap and fuzzy sphere results on the three operators. Note that for the last two operators the fuzzy sphere results do not identify them as conformal primary or descendant \cite{Zhou2024FuzzySphereDQCP}. Assume they are conformal primary operators, then their scaling dimensions are consistent with the bootstrap extremal solutions up to certain numerical errors.
The bootstrap results indicate the second lowest $SO(5)$ singlet scalar is near $4.8$ or smaller. With much larger $\Delta_{s'}>5.5$  the bootstrap cone near the QMC results will be excluded,  although its OPE coefficient is rather small. It would be interesting to resolve this discrepancy in the future studies.

\begin{table}[h]
\centering
\renewcommand{\arraystretch}{1.2}
\begin{tabular}{llll}
\toprule
\hline
Rep. & $\mathbf{1}$  &  $\mathbf{1}$ & $\mathbf{14}$   \\
\midrule[0.5pt] 
Spin & $0$ & $2$ & 2 \\
\midrule[0.5pt] 
Coef. & 0.008 & 0.045 & 0.091 \\
\midrule[0.5pt] 
F.S.
&
$>5.5$
&
$4.919^*$
&
$4.887^*$
\\
C.B.
&
$\sim 4.8$
&
4.259
&
4.591 
\\
\hline
\bottomrule
\end{tabular}
\caption{Operators with small OPE coefficients $\lambda^2<10^{-1}$.}
\label{tab:mismatch}
\end{table}

\section{Parameters and resources}

The parameters used in the bootstrap computations are 
\begin{equation}
   \Lambda=27:~ \textbf{pole}=20, ~\textbf{order}=80, ~\textbf{spins}=S_{27}, ~ \textbf{precision}=765,
\end{equation}
in which the set of spins $S_{27}$ is
\begin{equation}
    S_{27} = \{0,\ldots,26\}\cup \{29,30,33,34,37,38,41,42,45,46,49,50\}.
\end{equation}

The bootstrap programs used in this work are available at
\begin{itemize}
    \item \href{https://github.com/davidsd/sdpb}{\texttt{https://github.com/davidsd/sdpb}} for the SDPB, 
    \item \href{https://gitlab.com/bootstrapcollaboration/simpleboot}{\texttt{https://gitlab.com/bootstrapcollaboration/simpleboot}} for the bootstrap computations in Mathematica.
\end{itemize}

\bibliographystyle{utphys.bst}
\bibliography{HtDQCP}

@article{senthil2004science,
  author  = {Senthil, T. and Vishwanath, Ashvin and Balents, Leon and Sachdev, Subir and Fisher, Matthew P. A.},
  title   = {Deconfined Quantum Critical Points},
  journal = {Science},
  volume  = {303},
  number  = {5663},
  pages   = {1490--1494},
  year    = {2004},
  doi     = {10.1126/science.1091806}
}

@article{senthil2004prb,
  author  = {Senthil, T. and Balents, Leon and Sachdev, Subir and Vishwanath, Ashvin and Fisher, Matthew P. A.},
  title   = {Quantum Criticality beyond the Landau--Ginzburg--Wilson Paradigm},
  journal = {Physical Review B},
  volume  = {70},
  pages   = {144407},
  year    = {2004},
  doi     = {10.1103/PhysRevB.70.144407},
  eprint  = {cond-mat/0312617},
  archivePrefix = {arXiv}
}

@article{seiberg2016,
  author  = {Seiberg, Nathan and Senthil, T. and Wang, Chong and Witten, Edward},
  title   = {A Duality Web in \(2+1\) Dimensions and Condensed Matter Physics},
  journal = {Annals of Physics},
  volume  = {374},
  pages   = {395--433},
  year    = {2016},
  doi     = {10.1016/j.aop.2016.08.007},
  eprint  = {1606.01989},
  archivePrefix = {arXiv},
  primaryClass  = {hep-th}
}

@article{wang2017,
  author  = {Wang, Chong and Nahum, Adam and Metlitski, Max A. and Xu, Cenke and Senthil, T.},
  title   = {Deconfined Quantum Critical Points: Symmetries and Dualities},
  journal = {Physical Review X},
  volume  = {7},
  pages   = {031051},
  year    = {2017},
  doi     = {10.1103/PhysRevX.7.031051},
  eprint  = {1703.02426},
  archivePrefix = {arXiv},
  primaryClass  = {cond-mat.str-el}
}

@article{tanaka2005,
  author  = {Tanaka, Akihiro and Hu, Xiao},
  title   = {Many-Body Spin Berry Phases Emerging from the \(\pi\)-Flux State: Competition between Antiferromagnetism and the Valence-Bond-Solid State},
  journal = {Physical Review Letters},
  volume  = {95},
  pages   = {036402},
  year    = {2005},
  doi     = {10.1103/PhysRevLett.95.036402},
  eprint  = {cond-mat/0501365},
  archivePrefix = {arXiv}
}

@article{senthil2006,
  author  = {Senthil, T. and Fisher, Matthew P. A.},
  title   = {Competing Orders, Nonlinear Sigma Models, and Topological Terms in Quantum Magnets},
  journal = {Physical Review B},
  volume  = {74},
  pages   = {064405},
  year    = {2006},
  doi     = {10.1103/PhysRevB.74.064405},
  eprint  = {cond-mat/0510459},
  archivePrefix = {arXiv}
}

@article{cui2023,
  author  = {Cui, Yi and Liu, Lu and Lin, Huihang and Wu, Kai-Hsin and Hong, Wenshan and Liu, Xuefei and Li, Cong and Hu, Ze and Xi, Ning and Li, Shiliang and Yu, Rong and Sandvik, Anders W. and Yu, Weiqiang},
  title   = {Proximate Deconfined Quantum Critical Point in SrCu\(_2\)(BO\(_3\))\(_2\)},
  journal = {Science},
  volume  = {380},
  number  = {6650},
  pages   = {1179--1184},
  year    = {2023},
  doi     = {10.1126/science.adc9487},
  eprint  = {2204.08133},
  archivePrefix = {arXiv},
  primaryClass  = {cond-mat.str-el}
}

@article{guo2025,
  author  = {Guo, Jing and Wang, Pengyu and Huang, Cheng and Chen, Bin-Bin and Hong, Wenshan and Cai, Shu and Zhao, Jinyu and Han, Jinyu and Chen, Xintian and Zhou, Yazhou and Li, Shiliang and Wu, Qi and Meng, Zi Yang and Sun, Liling},
  title   = {Deconfined Quantum Critical Point Lost in Pressurized SrCu\(_2\)(BO\(_3\))\(_2\)},
  journal = {Communications Physics},
  volume  = {8},
  pages   = {72},
  year    = {2025},
  doi     = {10.1038/s42005-025-01976-8},
  eprint  = {2310.20128},
  archivePrefix = {arXiv},
  primaryClass  = {cond-mat.str-el}
}

@article{Zhu:2022gjc,
    author = "Zhu, Wei and Han, Chao and Huffman, Emilie and Hofmann, Johannes S. and He, Yin-Chen",
    title = "{Uncovering Conformal Symmetry in the 3D Ising Transition: State-Operator Correspondence from a Quantum Fuzzy Sphere Regularization}",
    eprint = "2210.13482",
    archivePrefix = "arXiv",
    primaryClass = "cond-mat.stat-mech",
    doi = "10.1103/PhysRevX.13.021009",
    journal = "Phys. Rev. X",
    volume = "13",
    number = "2",
    pages = "021009",
    year = "2023"
}

@article{Hu:2023xak,
    author = "Hu, Liangdong and He, Yin-Chen and Zhu, W.",
    title = "{Operator Product Expansion Coefficients of the 3D Ising Criticality via Quantum Fuzzy Spheres}",
    eprint = "2303.08844",
    archivePrefix = "arXiv",
    primaryClass = "cond-mat.stat-mech",
    doi = "10.1103/PhysRevLett.131.031601",
    journal = "Phys. Rev. Lett.",
    volume = "131",
    number = "3",
    pages = "031601",
    year = "2023"
}

@article{Li:2018lyb,
    author = "Li, Zhijin",
    title = "{Bootstrapping conformal QED$_{3}$ and deconfined quantum critical point}",
    eprint = "1812.09281",
    archivePrefix = "arXiv",
    primaryClass = "hep-th",
    doi = "10.1007/JHEP11(2022)005",
    journal = "JHEP",
    volume = "11",
    pages = "005",
    year = "2022"
}

@article{Chester:2023njo,
    author = "Chester, Shai M. and Su, Ning",
    title = "{Bootstrapping Deconfined Quantum Tricriticality}",
    eprint = "2310.08343",
    archivePrefix = "arXiv",
    primaryClass = "hep-th",
    doi = "10.1103/PhysRevLett.132.111601",
    journal = "Phys. Rev. Lett.",
    volume = "132",
    number = "11",
    pages = "111601",
    year = "2024"
}

@article{Nakayama:2016jhq,
    author = "Nakayama, Yu and Ohtsuki, Tomoki",
    title = "{Conformal Bootstrap Dashing Hopes of Emergent Symmetry}",
    eprint = "1602.07295",
    archivePrefix = "arXiv",
    primaryClass = "cond-mat.str-el",
    reportNumber = "IPMU16-0025",
    doi = "10.1103/PhysRevLett.117.131601",
    journal = "Phys. Rev. Lett.",
    volume = "117",
    number = "13",
    pages = "131601",
    year = "2016"
}

@article{He:2020azu,
    author = "He, Yin-Chen and Rong, Junchen and Su, Ning",
    title = "{Non-Wilson-Fisher kinks of $O(N)$ numerical bootstrap: from the deconfined phase transition to a putative new family of CFTs}",
    eprint = "2005.04250",
    archivePrefix = "arXiv",
    primaryClass = "hep-th",
    reportNumber = "DESY-20-224",
    doi = "10.21468/SciPostPhys.10.5.115",
    journal = "SciPost Phys.",
    volume = "10",
    number = "5",
    pages = "115",
    year = "2021"
}

@article{Kos:2013tga,
    author = "Kos, Filip and Poland, David and Simmons-Duffin, David",
    title = "{Bootstrapping the $O(N)$ vector models}",
    eprint = "1307.6856",
    archivePrefix = "arXiv",
    primaryClass = "hep-th",
    doi = "10.1007/JHEP06(2014)091",
    journal = "JHEP",
    volume = "06",
    pages = "091",
    year = "2014"
}

@article{Kos:2015mba,
    author = "Kos, Filip and Poland, David and Simmons-Duffin, David and Vichi, Alessandro",
    title = "{Bootstrapping the O(N) Archipelago}",
    eprint = "1504.07997",
    archivePrefix = "arXiv",
    primaryClass = "hep-th",
    reportNumber = "CERN-PH-TH-2015-097",
    doi = "10.1007/JHEP11(2015)106",
    journal = "JHEP",
    volume = "11",
    pages = "106",
    year = "2015"
}

@misc{cui2025,
      title={Two Plaquette-Singlet Phases and Emergent SO(5) Deconfined Quantum Criticality in SrCu2(BO3)2}, 
      author={Yi Cui and Kefan Du and Zhanlong Wu and Shuo Li and Pengtao Yang and Ying Chen and Xiaoyu Xu and Hongyu Chen and Chengchen Li and Juanjuan Liu and Bosen Wang and Wenshan Hong and Shiliang Li and Zhiyuan Xie and Jinguang Cheng and Bruce Normand and Rong Yu and Weiqiang Yu},
      year={2025},
      eprint={2411.00302},
      archivePrefix={arXiv},
      primaryClass={cond-mat.str-el},
      url={https://arxiv.org/abs/2411.00302}, 
}

@article{Chester:2025uxb,
    author = "Chester, Shai M. and Piazza, Alessandro and Reehorst, Marten and Su, Ning",
    title = "{Bootstrapping the simplest deconfined quantum critical point}",
    eprint = "2507.06283",
    archivePrefix = "arXiv",
    primaryClass = "hep-th",
    doi = "10.1103/ym81-bksd",
    journal = "Phys. Rev. D",
    volume = "113",
    number = "8",
    pages = "L081701",
    year = "2026"
}

@article{Li:2021emd,
    author = "Li, Zhijin",
    title = "{Conformality and self-duality of Nf=2 QED3}",
    eprint = "2107.09020",
    archivePrefix = "arXiv",
    primaryClass = "hep-th",
    doi = "10.1016/j.physletb.2022.137192",
    journal = "Phys. Lett. B",
    volume = "831",
    pages = "137192",
    year = "2022"
}

@article{Song:2023wlg,
    author = "Song, Menghan and Zhao, Jiarui and Cheng, Meng and Xu, Cenke and Scherer, Michael M. and Janssen, Lukas and Yang Meng, Zi",
    title = "{Evolution of entanglement entropy at SU(N) deconfined quantum critical points}",
    eprint = "2307.02547",
    archivePrefix = "arXiv",
    primaryClass = "cond-mat.str-el",
    doi = "10.1126/sciadv.adr0634",
    journal = "Sci. Adv.",
    volume = "11",
    number = "6",
    pages = "adr0634",
    year = "2025"
}

@article{Senthil:2018cru,
    author = "Senthil, T. and Son, Dam Thanh and Wang, Chong and Xu, Cenke",
    title = "{Duality between $(2+1)d$ Quantum Critical Points}",
    eprint = "1810.05174",
    archivePrefix = "arXiv",
    primaryClass = "cond-mat.str-el",
    doi = "10.1016/j.physrep.2019.09.001",
    journal = "Phys. Rept.",
    volume = "827",
    pages = "1--48",
    year = "2019"
}

@article{Zayed2017Plaquette,
  title        = {4-spin plaquette singlet state in the Shastry--Sutherland compound {SrCu$_2$(BO$_3$)$_2$}},
  author       = {Zayed, M. E. and R{\"u}egg, Ch. and Larrea J., J. and L{\"a}uchli, A. M. and Panagopoulos, C. and Saxena, S. S. and Ellerby, M. and McMorrow, D. F. and Str{\"a}ssle, Th. and Klotz, S. and Hamel, G. and Sadykov, R. A. and Pomjakushin, V. and Boehm, M. and Jim{\'e}nez-Ruiz, M. and Schneidewind, A. and Pomjakushina, E. and Stingaciu, M. and Conder, K. and R{\o}nnow, H. M.},
  journal      = {Nature Physics},
  volume       = {13},
  pages        = {962--966},
  year         = {2017},
  doi          = {10.1038/nphys4190}
}

@article{Guo2020HighPressureThermodynamics,
  title        = {Quantum phases of {SrCu$_2$(BO$_3$)$_2$} from high-pressure thermodynamics},
  author       = {Guo, Jing and Sun, Guangyu and Zhao, Bowen and Wang, Ling and Hong, Wenshan and Sidorov, Vladimir A. and Ma, Ning and Wu, Qi and Li, Shiliang and Meng, Zi Yang and Sandvik, Anders W. and Sun, Liling},
  journal      = {Physical Review Letters},
  volume       = {124},
  number       = {20},
  pages        = {206602},
  year         = {2020},
  doi          = {10.1103/PhysRevLett.124.206602}
}

@article{Shi2022ExtremeFieldPressure,
  title        = {Discovery of quantum phases in the Shastry--Sutherland compound {SrCu$_2$(BO$_3$)$_2$} under extreme conditions of field and pressure},
  author       = {Shi, Zhenzhong and Dissanayake, Sachith and Corboz, Philippe and Steinhardt, William and Graf, David and Silevitch, D. M. and Dabkowska, Hanna A. and Rosenbaum, T. F. and Mila, Fr{\'e}d{\'e}ric and Haravifard, Sara},
  journal      = {Nature Communications},
  volume       = {13},
  number       = {1},
  pages        = {2301},
  year         = {2022},
  doi          = {10.1038/s41467-022-30036-w}
}

@article{Lee2019ShastrySutherlandDQCP,
  title        = {Signatures of a deconfined phase transition on the Shastry--Sutherland lattice: Applications to quantum critical {SrCu$_2$(BO$_3$)$_2$}},
  author       = {Lee, Jong Yeon and You, Yi-Zhuang and Sachdev, Subir and Vishwanath, Ashvin},
  journal      = {Physical Review X},
  volume       = {9},
  number       = {4},
  pages        = {041037},
  year         = {2019},
  doi          = {10.1103/PhysRevX.9.041037}
}

@misc{Lin2024TwoPlaquetteSO5,
  title         = {Two plaquette-singlet phases and emergent {SO(5)} deconfined quantum criticality in {SrCu$_2$(BO$_3$)$_2$}},
  author        = {Lin, Huihang and Liu, Lu and Cui, Yi and Wu, Kai-Hsin and Hong, Wenshan and Liu, Xuefei and Li, Cong and Hu, Ze and Xi, Ning and Li, Shiliang and Yu, Rong and Sandvik, Anders W. and Yu, Weiqiang},
  year          = {2024},
  eprint        = {2411.00302},
  archivePrefix = {arXiv},
  primaryClass  = {cond-mat.str-el},
  doi           = {10.48550/arXiv.2411.00302}
}

@misc{senthil2023,
      title={Deconfined quantum critical points: a review}, 
      author={T. Senthil},
      year={2023},
      eprint={2306.12638},
      archivePrefix={arXiv},
      primaryClass={cond-mat.str-el},
      url={https://arxiv.org/abs/2306.12638}, 
}

@article{Nahum2015EmergentSO5,
  author  = {Nahum, Adam and Serna, Pablo and Chalker, J. T. and Ortu{\~n}o, M. and Somoza, A. M.},
  title   = {Emergent {SO(5)} Symmetry at the {N{\'e}el} to Valence-Bond-Solid Transition},
  journal = {Physical Review Letters},
  volume  = {115},
  pages   = {267203},
  year    = {2015},
  doi     = {10.1103/PhysRevLett.115.267203}
}

@article{Motrunich2004EmergentPhotons,
  author  = {Motrunich, O. I. and Vishwanath, Ashvin},
  title   = {Emergent Photons and Transitions in the {O(3)} Sigma Model with Hedgehog Suppression},
  journal = {Physical Review B},
  volume  = {70},
  pages   = {075104},
  year    = {2004},
  doi     = {10.1103/PhysRevB.70.075104}
}

@article{Sandvik2007EvidenceDQCP,
  author  = {Sandvik, Anders W.},
  title   = {Evidence for Deconfined Quantum Criticality in a Two-Dimensional {Heisenberg} Model with Four-Spin Interactions},
  journal = {Physical Review Letters},
  volume  = {98},
  pages   = {227202},
  year    = {2007},
  doi     = {10.1103/PhysRevLett.98.227202}
}

@article{Melko2008ScalingFan,
  author  = {Melko, Roger G. and Kaul, Ribhu K.},
  title   = {Scaling in the Fan of an Unconventional Quantum Critical Point},
  journal = {Physical Review Letters},
  volume  = {100},
  pages   = {017203},
  year    = {2008},
  doi     = {10.1103/PhysRevLett.100.017203}
}

@article{Jiang2008FirstOrderAFMVBS,
  author  = {Jiang, F.-J. and Nyfeler, M. and Chandrasekharan, S. and Wiese, U.-J.},
  title   = {From an Antiferromagnet to a Valence Bond Solid: Evidence for a First-Order Phase Transition},
  journal = {Journal of Statistical Mechanics: Theory and Experiment},
  volume  = {2008},
  pages   = {P02009},
  year    = {2008},
  doi     = {10.1088/1742-5468/2008/02/P02009}
}

@article{Kuklov2008GenericFirstOrder,
  author  = {Kuklov, A. B. and Matsumoto, M. and Prokof'ev, N. V. and Svistunov, B. V. and Troyer, M.},
  title   = {Deconfined Criticality: Generic First-Order Transition in the {SU(2)} Symmetry Case},
  journal = {Physical Review Letters},
  volume  = {101},
  pages   = {050405},
  year    = {2008},
  doi     = {10.1103/PhysRevLett.101.050405}
}

@article{Sandvik2010ContinuousTransition,
  author  = {Sandvik, Anders W.},
  title   = {Continuous Quantum Phase Transition Between an Antiferromagnet and a Valence-Bond Solid in Two Dimensions: Evidence for Logarithmic Corrections to Scaling},
  journal = {Physical Review Letters},
  volume  = {104},
  pages   = {177201},
  year    = {2010},
  doi     = {10.1103/PhysRevLett.104.177201}
}

@article{Lou2009SUNHeisenbergVBS,
  author  = {Lou, Jie and Sandvik, Anders W. and Kawashima, Naoki},
  title   = {Antiferromagnetic to Valence-Bond-Solid Transitions in Two-Dimensional {SU(N)} Heisenberg Models with Multispin Interactions},
  journal = {Physical Review B},
  volume  = {80},
  pages   = {180414},
  year    = {2009},
  doi     = {10.1103/PhysRevB.80.180414}
}

@article{Banerjee2010ImpurityTexture,
  author  = {Banerjee, Arnab and Damle, Kedar and Alet, Fabien},
  title   = {Impurity Spin Texture at a Deconfined Quantum Critical Point},
  journal = {Physical Review B},
  volume  = {82},
  pages   = {155139},
  year    = {2010},
  doi     = {10.1103/PhysRevB.82.155139}
}

@article{Kaul2012LargeNSUN,
  author  = {Kaul, Ribhu K. and Sandvik, Anders W.},
  title   = {Lattice Model for the {SU(N)} {N{\'e}el} to Valence-Bond Solid Quantum Phase Transition at Large {N}},
  journal = {Physical Review Letters},
  volume  = {108},
  pages   = {137201},
  year    = {2012},
  doi     = {10.1103/PhysRevLett.108.137201}
}

@article{Kaul2012BilayerSUN,
  author  = {Kaul, Ribhu K.},
  title   = {Quantum Phase Transitions in Bilayer {SU(N)} Antiferromagnets},
  journal = {Physical Review B},
  volume  = {85},
  pages   = {180411},
  year    = {2012},
  doi     = {10.1103/PhysRevB.85.180411}
}

@article{Bartosch2013CorrectionsScaling,
  author  = {Bartosch, Lorenz},
  title   = {Corrections to Scaling in the Critical Theory of Deconfined Criticality},
  journal = {Physical Review B},
  volume  = {88},
  pages   = {195140},
  year    = {2013},
  doi     = {10.1103/PhysRevB.88.195140}
}

@article{Chen2013DeconfinedFlow,
  author  = {Chen, Kun and Huang, Yuan and Deng, Youjin and Kuklov, A. B. and Prokof'ev, N. V. and Svistunov, B. V.},
  title   = {Deconfined Criticality Flow in the {Heisenberg} Model with Ring-Exchange Interactions},
  journal = {Physical Review Letters},
  volume  = {110},
  pages   = {185701},
  year    = {2013},
  doi     = {10.1103/PhysRevLett.110.185701}
}

@article{Pujari2013HoneycombDQCP,
  author  = {Pujari, Sumiran and Damle, Kedar and Alet, Fabien},
  title   = {{N{\'e}el}-State to Valence-Bond-Solid Transition on the Honeycomb Lattice: Evidence for Deconfined Criticality},
  journal = {Physical Review Letters},
  volume  = {111},
  pages   = {087203},
  year    = {2013},
  doi     = {10.1103/PhysRevLett.111.087203}
}

@article{Nahum2015ScalingViolations,
  author  = {Nahum, Adam and Chalker, J. T. and Serna, Pablo and Ortu{\~n}o, M. and Somoza, A. M.},
  title   = {Deconfined Quantum Criticality, Scaling Violations, and Classical Loop Models},
  journal = {Physical Review X},
  volume  = {5},
  pages   = {041048},
  year    = {2015},
  doi     = {10.1103/PhysRevX.5.041048}
}

@article{Sreejith2015HigherChargeMonopoles,
  author  = {Sreejith, G. J. and Powell, Stephen},
  title   = {Scaling Dimensions of Higher-Charge Monopoles at Deconfined Critical Points},
  journal = {Physical Review B},
  volume  = {92},
  pages   = {184413},
  year    = {2015},
  doi     = {10.1103/PhysRevB.92.184413}
}

@article{Sreejith2014CubicDimerMonomer,
  author  = {Sreejith, G. J. and Powell, Stephen},
  title   = {Critical Behavior in the Cubic Dimer Model at Nonzero Monomer Density},
  journal = {Physical Review B},
  volume  = {89},
  pages   = {014404},
  year    = {2014},
  doi     = {10.1103/PhysRevB.89.014404}
}

@article{Shao2016TwoLengthScales,
  author  = {Shao, Hui and Guo, Wenan and Sandvik, Anders W.},
  title   = {Quantum Criticality with Two Length Scales},
  journal = {Science},
  volume  = {352},
  pages   = {213--216},
  year    = {2016},
  doi     = {10.1126/science.aad5007}
}

@article{Sato2017DiracCompetingOrders,
  author  = {Sato, Toshihiro and Hohenadler, Martin and Assaad, Fakher F.},
  title   = {Dirac Fermions with Competing Orders: Non-{Landau} Transition with Emergent Symmetry},
  journal = {Physical Review Letters},
  volume  = {119},
  pages   = {197203},
  year    = {2017},
  doi     = {10.1103/PhysRevLett.119.197203}
}

@article{Serna2019ApproximateO4,
  author  = {Serna, Pablo and Nahum, Adam},
  title   = {Emergence and Spontaneous Breaking of Approximate {O(4)} Symmetry at a Weakly First-Order Deconfined Phase Transition},
  journal = {Physical Review B},
  volume  = {99},
  pages   = {195110},
  year    = {2019},
  doi     = {10.1103/PhysRevB.99.195110}
}

@article{Liu2019QSHSC,
  author  = {Liu, Yuhai and Wang, Zhenjiu and Sato, Toshihiro and Hohenadler, Martin and Wang, Chong and Guo, Wenan and Assaad, Fakher F.},
  title   = {Superconductivity from the Condensation of Topological Defects in a Quantum Spin-{Hall} Insulator},
  journal = {Nature Communications},
  volume  = {10},
  pages   = {2658},
  year    = {2019},
  doi     = {10.1038/s41467-019-10372-0}
}

@article{Wang2021SO5NLSM,
  author  = {Wang, Zhenjiu and Zaletel, Michael P. and Mong, Roger S. K. and Assaad, Fakher F.},
  title   = {Phases of the {(2+1)} Dimensional {SO(5)} Nonlinear Sigma Model with Topological Term},
  journal = {Physical Review Letters},
  volume  = {126},
  pages   = {045701},
  year    = {2021},
  doi     = {10.1103/PhysRevLett.126.045701}
}

@article{Liu2022GaplessQSL,
  title = {Emergence of Gapless Quantum Spin Liquid from Deconfined Quantum Critical Point},
  author = {Liu, Wen-Yuan and Hasik, Juraj and Gong, Shou-Shu and Poilblanc, Didier and Chen, Wei-Qiang and Gu, Zheng-Cheng},
  journal = {Physical Review X},
  volume = {12},
  pages = {031039},
  year = {2022},
  doi = {10.1103/PhysRevX.12.031039}
}

@article{Liu2022GaplessQSLGlobal, author = {Liu, Wen-Yuan and Gong, Shou-Shu and Li, Yu-Bin and Poilblanc, Didier and Chen, Wei-Qiang and Gu, Zheng-Cheng}, title = {Gapless quantum spin liquid and global phase diagram of the spin-1/2 {$J_1$--$J_2$} square antiferromagnetic {Heisenberg} model}, journal = {Science Bulletin}, volume = {67}, pages = {1034--1041}, year = {2022}, doi = {10.1016/j.scib.2022.03.010} }

@article{Zhao:2022dqcpEE,
  title         = {Scaling of Entanglement Entropy at Deconfined Quantum Criticality},
  author        = {Zhao, Jiarui and Wang, Yan-Cheng and Yan, Zheng and Cheng, Meng and Meng, Zi Yang},
  journal       = {Physical Review Letters},
  volume        = {128},
  number        = {1},
  pages         = {010601},
  year          = {2022},
  doi           = {10.1103/PhysRevLett.128.010601},
  eprint        = {2107.06305},
  archivePrefix = {arXiv},
  primaryClass  = {cond-mat.str-el}
}

@article{Zhou2024FuzzySphereDQCP,
  title = {The $\mathrm{SO}(5)$ Deconfined Phase Transition under the Fuzzy-Sphere Microscope: Approximate Conformal Symmetry, Pseudocriticality, and Operator Spectrum},
  author = {Zhou, Zheng and Hu, Liangdong and Zhu, W. and He, Yin-Chen},
  journal = {Physical Review X},
  volume = {14},
  pages = {021044},
  year = {2024},
  doi = {10.1103/PhysRevX.14.021044},
  eprint = {2306.16435},
  archivePrefix = {arXiv},
  primaryClass = {cond-mat.str-el}
}

@article{Chen2024SO5NLSMSphere,
  title = {Phases of $(2+1)$D $\mathrm{SO}(5)$ Nonlinear Sigma Model with a Topological Term on a Sphere: Multicritical Point and Disorder Phase},
  author = {Chen, Bin-Bin and Zhang, Xiangyu and Wang, Yan-Cheng and Sun, Kai and Meng, Zi Yang},
  journal = {Physical Review Letters},
  volume = {132},
  pages = {246503},
  year = {2024},
  doi = {10.1103/PhysRevLett.132.246503},
  eprint = {2307.05307},
  archivePrefix = {arXiv},
  primaryClass = {cond-mat.str-el}
}

@article{Chen2024EmergentConformalSO5,
  title = {Emergent Conformal Symmetry at the Multicritical Point of $(2+1)$D $\mathrm{SO}(5)$ Model with Wess-Zumino-Witten Term on Sphere},
  author = {Chen, Bin-Bin and Zhang, Xiangyu and Meng, Zi Yang},
  journal = {Physical Review B},
  volume = {110},
  pages = {125153},
  year = {2024},
  doi = {10.1103/PhysRevB.110.125153},
  eprint = {2405.04470},
  archivePrefix = {arXiv},
  primaryClass = {cond-mat.str-el}
}

@article{Song2025EvolutionEE,
  title = {Evolution of Entanglement Entropy at $\mathrm{SU}(N)$ Deconfined Quantum Critical Points},
  author = {Song, Menghan and Zhao, Jiarui and Cheng, Meng and Xu, Cenke and Scherer, Michael M. and Janssen, Lukas and Meng, Zi Yang},
  journal = {Science Advances},
  volume = {11},
  number = {6},
  pages = {eadr0634},
  year = {2025},
  doi = {10.1126/sciadv.adr0634},
  eprint = {2307.02547},
  archivePrefix = {arXiv},
  primaryClass = {cond-mat.str-el}
}

@article{Deng2024DiagnosingEE,
  title = {Diagnosing Quantum Phase Transition Order and Deconfined Criticality via Entanglement Entropy},
  author = {Deng, Zehui and Liu, Lu and Guo, Wenan and Lin, H. Q.},
  journal = {Physical Review Letters},
  volume = {133},
  pages = {100402},
  year = {2024},
  doi = {10.1103/PhysRevLett.133.100402},
  eprint = {2401.12838},
  archivePrefix = {arXiv},
  primaryClass = {cond-mat.str-el}
}

@article{DEmidio2024EntanglementSO5,
  title = {Entanglement Entropy and Deconfined Criticality: Emergent $\mathrm{SO}(5)$ Symmetry and Proper Lattice Bipartition},
  author = {D'Emidio, Jonathan and Sandvik, Anders W.},
  journal = {Physical Review Letters},
  volume = {133},
  pages = {166702},
  year = {2024},
  doi = {10.1103/PhysRevLett.133.166702},
  eprint = {2401.14396},
  archivePrefix = {arXiv},
  primaryClass = {cond-mat.str-el}
}

@misc{Takahashi2024,
  title         = {$SO(5)$ Multicriticality in Two-Dimensional Quantum Magnets},
  author        = {Takahashi, Jun and Shao, Hui and Zhao, Bowen and Guo, Wenan and Sandvik, Anders W.},
  year          = {2024},
  eprint        = {2405.06607},
  archivePrefix = {arXiv},
  primaryClass  = {cond-mat.str-el}
}

@article{Zhu:2026Bi,
  title         = {Bipartite Entanglement and Surface Criticality: The Extra Contribution of the Nonordinary Edge in Entanglement},
  author        = {Zhu, Yanzhang and Liu, Zenan and Wang, Zhe and Wang, Yan-Cheng and Yan, Zheng},
  journal       = {Physical Review Letters},
  volume        = {136},
  pages         = {046501},
  year          = {2026},
  doi           = {10.1103/tkx5-kzhh},
  eprint        = {2508.07277},
  archivePrefix = {arXiv},
  primaryClass  = {cond-mat.str-el}
}

@misc{daliao2026,
      title={Numerical evidence of a critical point in the (2+1)D SO(5) nonlinear sigma model with Wess-Zumino-Witten term}, 
      author={Yuan Da Liao and Bin-Bin Chen and Fakher F. Assaad and Lukas Janssen and Zi Yang Meng},
      year={2026},
      eprint={2605.03700},
      archivePrefix={arXiv},
      primaryClass={cond-mat.str-el},
      url={https://arxiv.org/abs/2605.03700}, 
}

@article{Poland2019,
  author        = {Poland, David and Rychkov, Slava and Vichi, Alessandro},
  title         = {The Conformal Bootstrap: Theory, Numerical Techniques, and Applications},
  journal       = {Reviews of Modern Physics},
  volume        = {91},
  pages         = {015002},
  year          = {2019},
  doi           = {10.1103/RevModPhys.91.015002},
  eprint        = {1805.04405},
  archivePrefix = {arXiv},
  primaryClass  = {hep-th}
}

@article{Gorbenko2018,
  title         = {Walking, Weak First-Order Transitions, and Complex CFTs},
  author        = {Gorbenko, Victor and Rychkov, Slava and Zan, Bernardo},
  journal       = {Journal of High Energy Physics},
  volume        = {2018},
  number        = {10},
  pages         = {108},
  year          = {2018},
  doi           = {10.1007/JHEP10(2018)108},
  eprint        = {1807.11512},
  archivePrefix = {arXiv},
  primaryClass  = {hep-th}
}

@article{Gorbenko2018II,
  title         = {Walking, Weak First-Order Transitions, and Complex CFTs II: Two-Dimensional Potts Model at $Q>4$},
  author        = {Gorbenko, Victor and Rychkov, Slava and Zan, Bernardo},
  journal       = {SciPost Physics},
  volume        = {5},
  pages         = {050},
  year          = {2018},
  doi           = {10.21468/SciPostPhys.5.5.050},
  eprint        = {1808.04380},
  archivePrefix = {arXiv},
  primaryClass  = {hep-th}
}

@article{Ma2020,
  title         = {Theory of Deconfined Pseudocriticality},
  author        = {Ma, Ruochen and Wang, Chong},
  journal       = {Physical Review B},
  volume        = {102},
  pages         = {020407},
  year          = {2020},
  doi           = {10.1103/PhysRevB.102.020407},
  eprint        = {1912.12315},
  archivePrefix = {arXiv},
  primaryClass  = {cond-mat.str-el}
}

@article{Zhao2020,
  title         = {Multicritical Deconfined Quantum Criticality and Lifshitz Point of a Helical Valence-Bond Phase},
  author        = {Zhao, Bowen and Takahashi, Jun and Sandvik, Anders W.},
  journal       = {Physical Review Letters},
  volume        = {125},
  pages         = {257204},
  year          = {2020},
  doi           = {10.1103/PhysRevLett.125.257204},
  eprint        = {2005.10184},
  archivePrefix = {arXiv},
  primaryClass  = {cond-mat.str-el}
}

@article{Rattazzi:2008pe,
  author        = {Rattazzi, Riccardo and Rychkov, Vyacheslav S. and Tonni, Erik and Vichi, Alessandro},
  title         = {Bounding scalar operator dimensions in 4D CFT},
  journal       = {JHEP},
  volume        = {12},
  pages         = {031},
  year          = {2008},
  doi           = {10.1088/1126-6708/2008/12/031},
  eprint        = {0807.0004},
  archivePrefix = {arXiv},
  primaryClass  = {hep-th}
}

@article{Rychkov:2023wsd,
  author        = {Rychkov, Slava and Su, Ning},
  title         = {New Developments in the Numerical Conformal Bootstrap},
  journal       = {Rev. Mod. Phys.},
  volume        = {96},
  number        = {4},
  pages         = {045004},
  year          = {2024},
  doi           = {10.1103/RevModPhys.96.045004},
  eprint        = {2311.15844},
  archivePrefix = {arXiv},
  primaryClass  = {hep-th}
}

@article{Poland:2016chs,
  author  = {Poland, David and Simmons-Duffin, David},
  title   = {The conformal bootstrap},
  journal = {Nature Phys.},
  volume  = {12},
  number  = {6},
  pages   = {535--539},
  year    = {2016},
  doi     = {10.1038/nphys3761}
}

@article{Zhang1997SO5,
  author  = {Zhang, Shou-Cheng},
  title   = {A unified theory based on {SO(5)} symmetry of superconductivity and antiferromagnetism},
  journal = {Science},
  volume  = {275},
  number  = {5303},
  pages   = {1089--1096},
  year    = {1997},
  doi     = {10.1126/science.275.5303.1089}
}

@article{Demler2004,
  author        = {Demler, Eugene and Hanke, Werner and Zhang, Shou-Cheng},
  title         = {{SO(5)} theory of antiferromagnetism and superconductivity},
  journal       = {Reviews of Modern Physics},
  volume        = {76},
  pages         = {909--974},
  year          = {2004},
  doi           = {10.1103/RevModPhys.76.909},
  eprint        = {cond-mat/0405038},
  archivePrefix = {arXiv},
  primaryClass  = {cond-mat.str-el}
}

@article{Calabrese2003,
  author        = {Calabrese, Pasquale and Pelissetto, Andrea and Vicari, Ettore},
  title         = {Multicritical phenomena in {$O(n_1)\oplus O(n_2)$}-symmetric theories},
  journal       = {Physical Review B},
  volume        = {67},
  pages         = {054505},
  year          = {2003},
  doi           = {10.1103/PhysRevB.67.054505},
  eprint        = {cond-mat/0209580},
  archivePrefix = {arXiv}
}

@article{Hasenbusch2005,
  author        = {Hasenbusch, Martin and Pelissetto, Andrea and Vicari, Ettore},
  title         = {Instability of the {$O(5)$} multicritical behavior in the {SO(5)} theory of high-{$T_c$} superconductors},
  journal       = {Physical Review B},
  volume        = {72},
  pages         = {014532},
  year          = {2005},
  doi           = {10.1103/PhysRevB.72.014532},
  eprint        = {cond-mat/0502327},
  archivePrefix = {arXiv}
}

@article{Hasenbusch2022,
  author        = {Hasenbusch, Martin},
  title         = {Three-dimensional {$O(N)$}-invariant {$\phi^4$} models at criticality for {$N\ge 4$}},
  journal       = {Physical Review B},
  volume        = {105},
  pages         = {054428},
  year          = {2022},
  doi           = {10.1103/PhysRevB.105.054428},
  eprint        = {2112.03783},
  archivePrefix = {arXiv},
  primaryClass  = {hep-lat}
}

@article{Bonati2025,
  author        = {Bonati, Claudio and Soler Calero, Ivan},
  title         = {{O(5)} multicriticality in the {3D} two flavor {SU(2)} lattice gauge {Higgs} model},
  journal       = {Physical Review E},
  volume        = {112},
  pages         = {024112},
  year          = {2025},
  doi           = {10.1103/l6wr-6p69},
  eprint        = {2505.03446},
  archivePrefix = {arXiv},
  primaryClass  = {hep-lat}
}

@article{Reehorst2021,
  author        = {Reehorst, Marten and Rychkov, Slava and Simmons-Duffin, David and Sirois, Benoit and Su, Ning and van Rees, Balt},
  title         = {Navigator Function for the Conformal Bootstrap},
  journal       = {SciPost Physics},
  volume        = {11},
  number        = {3},
  pages         = {072},
  year          = {2021},
  doi           = {10.21468/SciPostPhys.11.3.072},
  eprint        = {2104.09518},
  archivePrefix = {arXiv},
  primaryClass  = {hep-th}
}

@article{Li:2020bnb,
    author = "Li, Zhijin and Poland, David",
    title = "{Searching for gauge theories with the conformal bootstrap}",
    eprint = "2005.01721",
    archivePrefix = "arXiv",
    primaryClass = "hep-th",
    doi = "10.1007/JHEP03(2021)172",
    journal = "JHEP",
    volume = "03",
    pages = "172",
    year = "2021"
}

@article{Li:2020tsm,
    author = "Li, Zhijin",
    title = "{Symmetries of conformal correlation functions}",
    eprint = "2006.05119",
    archivePrefix = "arXiv",
    primaryClass = "hep-th",
    doi = "10.1103/PhysRevD.105.085018",
    journal = "Phys. Rev. D",
    volume = "105",
    number = "8",
    pages = "085018",
    year = "2022"
}

@article{Albayrak:2021xtd,
    author = "Albayrak, Soner and Erramilli, Rajeev S. and Li, Zhijin and Poland, David and Xin, Yuan",
    title = "{Bootstrapping $N_f$=4 conformal QED$_3$}",
    eprint = "2112.02106",
    archivePrefix = "arXiv",
    primaryClass = "hep-th",
    doi = "10.1103/PhysRevD.105.085008",
    journal = "Phys. Rev. D",
    volume = "105",
    number = "8",
    pages = "085008",
    year = "2022"
}

@article{ZhaoWeinbergSandvik2019O4,
  title        = {Symmetry-enhanced discontinuous phase transition in a two-dimensional quantum magnet},
  author       = {Zhao, Bowen and Weinberg, Phillip and Sandvik, Anders W.},
  journal      = {Nature Physics},
  volume       = {15},
  pages        = {678--682},
  year         = {2019},
  doi          = {10.1038/s41567-019-0484-x}
}

@article{Qin:2017xqg,
  author        = {Qin, Yan Qi and He, Yuan-Yao and You, Yi-Zhuang and Lu, Zhong-Yi and Sen, Arnab and Sandvik, Anders W. and Xu, Cenke and Meng, Zi Yang},
  title         = {Duality between the deconfined quantum-critical point and the bosonic topological transition},
  journal       = {Phys. Rev. X},
  volume        = {7},
  number        = {3},
  pages         = {031052},
  year          = {2017},
  doi           = {10.1103/PhysRevX.7.031052},
  eprint        = {1705.10670},
  archivePrefix = {arXiv},
  primaryClass  = {cond-mat.str-el}
}

@article{Liu2024EmergentSymmetry,
  title = {Emergent symmetry in quantum phase transition: From deconfined quantum critical point to gapless quantum spin liquid},
  author = {Liu, Wen-Yuan and Gong, Shou-Shu and Chen, Wei-Qiang and Gu, Zheng-Cheng},
  journal = {Science Bulletin},
  volume = {69},
  number = {2},
  pages = {190--196},
  year = {2024},
  doi = {10.1016/j.scib.2023.11.057},
  eprint = {2212.00707},
  archivePrefix = {arXiv},
  primaryClass = {cond-mat.str-el}
}

@article{ShastrySutherland1981,
  author  = {Shastry, B. S. and Sutherland, Bill},
  title   = {Exact ground state of a quantum mechanical antiferromagnet},
  journal = {Physica B+C},
  volume  = {108},
  pages   = {1069--1070},
  year    = {1981},
  doi     = {10.1016/0378-4363(81)90838-X}
}

@misc{li2024,
      title={Conformality loss and short-range crossover in long-range conformal field theories}, 
      author={Zhijin Li},
      year={2024},
      eprint={2409.19392},
      archivePrefix={arXiv},
      primaryClass={hep-th},
      url={https://arxiv.org/abs/2409.19392}, 
}

@misc{YangYaoSandvik2020LongRangeDQCP,
  author        = {Yang, Sibin and Yao, Dao-Xin and Sandvik, Anders W.},
  title         = {Deconfined quantum criticality in spin-1/2 chains with long-range interactions},
  year          = {2020},
  eprint        = {2001.02821},
  archivePrefix = {arXiv},
  primaryClass  = {physics.comp-ph}
}

@article{Jian2021NonlocalNeelVBS,
  author  = {Jian, Chao-Ming and Xu, Yichen and Wu, Xiao-Chuan and Xu, Cenke},
  title   = {Continuous N{\'e}el-VBS quantum phase transition in non-local one-dimensional systems with SO(3) symmetry},
  journal = {SciPost Physics},
  volume  = {10},
  pages   = {033},
  year    = {2021},
  doi     = {10.21468/SciPostPhys.10.2.033},
  eprint  = {2004.07852},
  archivePrefix = {arXiv},
  primaryClass = {cond-mat.str-el}
}

@article{Xu2022NonlocalDQCP,
  author  = {Xu, Yichen and Wu, Xiao-Chuan and Jian, Chao-Ming and Xu, Cenke},
  title   = {Deconfined quantum critical point with non-locality},
  journal = {Physical Review B},
  volume  = {106},
  pages   = {155131},
  year    = {2022},
  doi     = {10.1103/PhysRevB.106.155131}
}

@article{Romen2024LongRangeAnisotropicHeisenberg,
  author  = {Romen, Anton and Birnkammer, Stefan and Knap, Michael},
  title   = {Deconfined Quantum Criticality in the long-range, anisotropic Heisenberg Chain},
  journal = {SciPost Physics Core},
  volume  = {7},
  pages   = {008},
  year    = {2024},
  doi     = {10.21468/SciPostPhysCore.7.1.008},
  eprint  = {2311.06350},
  archivePrefix = {arXiv},
  primaryClass = {cond-mat.str-el}
}

@article{Calabrese2002,
  title = {Critical structure factors of bilinear fields in $\mathrm{O}(N)$ vector models},
  author = {Calabrese, Pasquale and Pelissetto, Andrea and Vicari, Ettore},
  journal = {Phys. Rev. E},
  volume = {65},
  issue = {4},
  pages = {046115},
  numpages = {16},
  year = {2002},
  month = {Apr},
  publisher = {American Physical Society},
  doi = {10.1103/PhysRevE.65.046115},
  url = {https://link.aps.org/doi/10.1103/PhysRevE.65.046115}
}

@article{Metlitski2008,
  title = {Monopoles in ${\text{CP}}^{N\ensuremath{-}1}$ model via the state-operator correspondence},
  author = {Metlitski, Max A. and Hermele, Michael and Senthil, T. and Fisher, Matthew P. A.},
  journal = {Phys. Rev. B},
  volume = {78},
  issue = {21},
  pages = {214418},
  numpages = {10},
  year = {2008},
  month = {Dec},
  publisher = {American Physical Society},
  doi = {10.1103/PhysRevB.78.214418},
  url = {https://link.aps.org/doi/10.1103/PhysRevB.78.214418}
}

@article{Dyer:2015zha,
    author = "Dyer, Ethan and Mezei, M{\'a}rk and Pufu, Silviu S. and Sachdev, Subir",
    title = "{Scaling dimensions of monopole operators in the $ \mathbb{C}{\mathrm{\mathbb{P}}}^{N_b-1} $ theory in 2 $+$ 1 dimensions}",
    eprint = "1504.00368",
    archivePrefix = "arXiv",
    primaryClass = "hep-th",
    reportNumber = "PUPT-2479",
    doi = "10.1007/JHEP03(2016)111",
    journal = "JHEP",
    volume = "06",
    pages = "037",
    year = "2015",
    note = "[Erratum: JHEP 03, 111 (2016)]"
}

@article{Chester:2024waw,
    author = "Chester, Shai M. and Komargodski, Zohar",
    title = "{Symmetry enhancement, symmetry-protected topological absorption, and duality in QED3}",
    eprint = "2409.17913",
    archivePrefix = "arXiv",
    primaryClass = "hep-th",
    doi = "10.1103/pgm4-dtlp",
    journal = "Phys. Rev. B",
    volume = "112",
    number = "4",
    pages = "L041113",
    year = "2025"
}

@misc{dumitrescu2024symmetrybreakingmonopolecondensation,
      title={Symmetry Breaking from Monopole Condensation in QED$_3$}, 
      author={Thomas T. Dumitrescu and Pierluigi Niro and Ryan Thorngren},
      year={2024},
      eprint={2410.05366},
      archivePrefix={arXiv},
      primaryClass={hep-th},
      url={https://arxiv.org/abs/2410.05366}, 
}

@article{Chester:2019ifh,
  author        = {Chester, Shai M. and Landry, Walter and Liu, Junyu
                   and Poland, David and Simmons-Duffin, David
                   and Su, Ning and Vichi, Alessandro},
  title         = {{Carving out OPE space and precise O(2) model critical exponents}},
  eprint        = {1912.03324},
  archivePrefix = {arXiv},
  primaryClass  = {hep-th},
  journal       = {JHEP},
  volume        = {06},
  pages         = {142},
  year          = {2020},
  doi           = {10.1007/JHEP06(2020)142}
}

@article{Chang:2024whu,
  author        = {Chang, Cyuan-Han and Dommes, Vasiliy and Erramilli, Rajeev S.
                   and Homrich, Alexandre and Kravchuk, Petr and Liu, Aike
                   and Mitchell, Matthew S. and Poland, David
                   and Simmons-Duffin, David},
  title         = {{Bootstrapping the 3d Ising Stress Tensor}},
  eprint        = {2411.15300},
  archivePrefix = {arXiv},
  primaryClass  = {hep-th},
  reportNumber  = {CALT-TH 2024-047},
  journal       = {JHEP},
  volume        = {03},
  pages         = {136},
  year          = {2025},
  doi           = {10.1007/JHEP03(2025)136}
}

@article{Chester:2020iyt,
  author        = {Chester, Shai M. and Landry, Walter and Liu, Junyu
                   and Poland, David and Simmons-Duffin, David
                   and Su, Ning and Vichi, Alessandro},
  title         = {{Bootstrapping Heisenberg Magnets and their Cubic Instability}},
  eprint        = {2011.14647},
  archivePrefix = {arXiv},
  primaryClass  = {hep-th},
  journal       = {Phys. Rev. D},
  volume        = {104},
  number        = {10},
  pages         = {105013},
  year          = {2021},
  doi           = {10.1103/PhysRevD.104.105013}
}

@article{Simmons-Duffin:2015qma,
  author        = {Simmons-Duffin, David},
  title         = {{A Semidefinite Program Solver for the Conformal Bootstrap}},
  eprint        = {1502.02033},
  archivePrefix = {arXiv},
  primaryClass  = {hep-th},
  journal       = {JHEP},
  volume        = {06},
  pages         = {174},
  year          = {2015},
  doi           = {10.1007/JHEP06(2015)174}
}

@article{Landry:2019qug,
  author        = {Landry, Walter and Simmons-Duffin, David},
  title         = {{Scaling the semidefinite program solver SDPB}},
  eprint        = {1909.09745},
  archivePrefix = {arXiv},
  primaryClass  = {hep-th},
  year          = {2019},
  doi           = {10.48550/arXiv.1909.09745}
}

@article{Kos:2014bka,
    author = "Kos, Filip and Poland, David and Simmons-Duffin, David",
    title = "{Bootstrapping Mixed Correlators in the 3D Ising Model}",
    eprint = "1406.4858",
    archivePrefix = "arXiv",
    primaryClass = "hep-th",
    doi = "10.1007/JHEP11(2014)109",
    journal = "JHEP",
    volume = "11",
    pages = "109",
    year = "2014"
}

@article{Liu:2020tpf,
  author        = {Liu, Junyu and Meltzer, David and Poland, David and Simmons-Duffin, David},
  title         = {{The Lorentzian inversion formula and the spectrum of the 3d O(2) CFT}},
  journal       = {JHEP},
  volume        = {09},
  pages         = {115},
  year          = {2020},
  eprint        = {2007.07914},
  archivePrefix = {arXiv},
  primaryClass  = {hep-th},
  doi           = {10.1007/JHEP09(2020)115},
  note          = {[Erratum: JHEP 01, 206 (2021)]}
}

@article{Yang:2026SO5O4Flow,
  title   = {Conformal Operator Flows of the Deconfined Quantum Criticality from {$\mathrm{SO}(5)$} to {$\mathrm{O}(4)$}},
  author  = {Yang, Shuai and Hu, Liang-dong and Han, Chao and Zhu, W. and Chen, Yan},
  journal = {Physical Review Letters},
  volume  = {136},
  pages   = {076505},
  year    = {2026},
  doi     = {10.1103/PhysRevLett.136.076505},
  eprint  = {2507.01322},
  archivePrefix = {arXiv},
  primaryClass  = {cond-mat.str-el}
}

@misc{zou2025O4,
      title={Unraveling Deconfined Quantum Criticality in Non-Hermitian Easy-Plane $J$-$Q$ Model}, 
      author={Xuan Zou and Shuai Yin and Zi-Xiang Li and Hong Yao},
      year={2025},
      eprint={2511.03456},
      archivePrefix={arXiv},
      primaryClass={cond-mat.str-el},
      url={https://arxiv.org/abs/2511.03456}, 
}
\end{document}